\documentclass[twocolumn,twocolappendix,tighten,times]{aastex631}

\usepackage{amsmath}
\usepackage{newtxmath}
\usepackage{multirow}

\received{July 25, 2022}
\revised{November 1, 2022}
\accepted{November 25, 2022}

\submitjournal{\apj}

\shorttitle{A Method for Discovering Reionization-era Lensed Quasars}
\shortauthors{Andika et al.}

\graphicspath{{./}{figures/}}

\begin{document}

\title{When Spectral Modeling Meets Convolutional Networks:\\ A Method for Discovering Reionization-era Lensed Quasars in Multi-band Imaging Data}

\correspondingauthor{Irham Taufik Andika}
\email{irham.andika@tum.de}

\author[0000-0001-6102-9526]{Irham Taufik Andika}
\affiliation{Max-Planck-Institut f\"{u}r Astronomie, K\"{o}nigstuhl 17, D-69117 Heidelberg, Germany}
\affiliation{Sterrenkundig Observatorium, Universiteit Gent, Krijgslaan 281 S9, B-9000 Gent, Belgium}
\affiliation{Physik-Department, Technische Universit\"{a}t M\"{u}nchen, James-Franck-Str. 1, D-85748 Garching bei M\"{u}nchen, Germany}

\author[0000-0003-3804-2137]{Knud Jahnke}
\affiliation{Max-Planck-Institut f\"{u}r Astronomie, K\"{o}nigstuhl 17, D-69117 Heidelberg, Germany}

\author[0000-0002-5027-0135]{Arjen van der Wel}
\affiliation{Sterrenkundig Observatorium, Universiteit Gent, Krijgslaan 281 S9, B-9000 Gent, Belgium}

\author[0000-0002-2931-7824]{Eduardo Ba\~{n}ados}
\affiliation{Max-Planck-Institut f\"{u}r Astronomie, K\"{o}nigstuhl 17, D-69117 Heidelberg, Germany}

\author[0000-0001-8582-7012]{Sarah E. I. Bosman}
\affiliation{Max-Planck-Institut f\"{u}r Astronomie, K\"{o}nigstuhl 17, D-69117 Heidelberg, Germany}

\author[0000-0003-0821-3644]{Frederick B. Davies}
\affiliation{Max-Planck-Institut f\"{u}r Astronomie, K\"{o}nigstuhl 17, D-69117 Heidelberg, Germany}

\author[0000-0003-2895-6218]{Anna-Christina Eilers}
\altaffiliation{NASA Hubble Fellow}
\affiliation{MIT Kavli Institute for Astrophysics and Space Research, 77 Massachusetts Ave., Cambridge, MA 02139, USA}

\author[0000-0001-6282-5778]{Anton Timur Jaelani}
\affiliation{Astronomy Research Group and Bosscha Observatory, FMIPA, Institut Teknologi Bandung, Jl. Ganesha 10, Bandung 40132, Indonesia}
\affiliation{U-CoE AI-VLB, Institut Teknologi Bandung, Jl. Ganesha 10, Bandung 40132, Indonesia}

\author[0000-0002-5941-5214]{Chiara Mazzucchelli}
\affiliation{N\'{u}cleo de Astronom\'{i}a, Facultad de Ingenier\'{i}a y Ciencias, Universidad Diego Portales, Av. Ej\'{e}rcito 441, Santiago, 8320000, Chile}

\author[0000-0003-2984-6803]{Masafusa Onoue}
\affiliation{Kavli Institute for Astronomy and Astrophysics, Peking University, Beijing 100871, China}
\affiliation{Kavli Institute for the Physics and Mathematics of the Universe (Kavli IPMU, WPI), The University of Tokyo, Chiba 277-8583, Japan}

\author[0000-0002-4544-8242]{Jan-Torge Schindler}
\affiliation{Max-Planck-Institut f\"{u}r Astronomie, K\"{o}nigstuhl 17, D-69117 Heidelberg, Germany}
\affiliation{Leiden Observatory, Leiden University, PO Box 9513, 2300 RA Leiden, The Netherlands}

\begin{abstract}
Over the last two decades, around 300 quasars have been discovered at $z\gtrsim6$, yet only one has identified as being strongly gravitationally lensed. 
We explore a new approach -- enlarging the permitted spectral parameter space, while introducing a new spatial geometry veto criterion -- which is implemented via image-based deep learning. 
We first apply this approach to a systematic search for reionization-era lensed quasars, using data from the Dark Energy Survey, the Visible and Infrared Survey Telescope for Astronomy Hemisphere Survey, and the Wide-field Infrared Survey Explorer.
Our search method consists of two main parts: 
(i) the preselection of the candidates based on their spectral energy distributions (SEDs) using catalog-level photometry and 
(ii) relative probabilities calculation of the candidates being a lens or some contaminant, utilizing a convolutional neural network (CNN) classification.
The training data sets are constructed by painting deflected point-source lights over actual galaxy images, to generate realistic galaxy-quasar lens models, optimized to find systems with small image separations, i.e., Einstein radii of $\theta_\mathrm{E} \leq 1\arcsec$.
Visual inspection is then performed for sources with CNN scores of $P_\mathrm{lens} > 0.1$, which leads us to obtain 36 newly selected lens candidates, which are awaiting spectroscopic confirmation.
These findings show that automated SED modeling and deep learning pipelines, supported by modest human input, are a promising route for detecting strong lenses from large catalogs that can overcome the veto limitations of primarily dropout-based SED selection approaches.
\end{abstract}


\section{Introduction} \label{sec:intro}

Quasars are among the brightest sources in the universe, powered by the accretion of matter onto supermassive black holes (SMBHs).
In the high-redshift frontier, these sources are prime laboratories for understanding the formation, growth, and structure of the first SMBHs and galaxies \citep{2020A&A...642L...1M,2022MNRAS.509.1885P}.
Key results from the last two decades of $z\gtrsim6$ quasar surveys include:
(i) The number density of distant quasars places stringent constraints on the processes necessary to seed and develop $>10^9~M_\odot$ black holes within the first billion years of cosmic history \citep{2020ARA&A..58...27I}.
(ii) High-$z$ quasars most commonly reside in massive star-forming host galaxies with a wide range of kinematics and large-scale environments \citep{2018ApJ...854...97D,2021ApJ...911..141N,2022ApJ...927..141M}.
(iii) Hydrogen absorption in the spectra of high-$z$ quasars indicates a largely neutral intergalactic medium at $z\gtrsim7$ \citep{2018Natur.553..473B,2020ApJ...897L..14Y,2021ApJ...907L...1W} and an end of cosmic reionization by $z\sim5.3$ \citep{2022MNRAS.514...55B}.

Previous high-$z$ quasar searches have been mainly focused on the most luminous, massive, and active SMBHs, while those fainter ones with modest to low accretion or lower mass are challenging to find \citep[e.g.,][]{2006AJ....132..117F,2010AJ....139..906W,2011Natur.474..616M,2015ApJ...801L..11V,2016ApJS..227...11B,2017ApJ...849...91M,2019MNRAS.487.1874R,2019AJ....157..236Y,2019ApJ...884...30W,2019MNRAS.484.5142P,2021AJ....162...72W,2021ApJ...909...80B,2022PhDT.........1A}.
This is due to selections having to filter out rare quasars from more frequent galaxies and stars, hence having to focus on the brightest population first.
Another reason is that many wide-sky surveys in the prior years are too shallow, so one is automatically restricted to studying the brightest population.
Some efforts have been made to have deeper wide-sky surveys \citep[e.g.,][]{2018A&A...617A.127P,2018ApJS..237....5M}, because exploring the fainter, currently mostly hidden, and lower mass population of high-$z$ quasars will provide knowledge on the quasar luminosity function, impacting our understanding of the quasar contributions to the universe's reionization \citep{2017ApJ...851...50M,2018ApJ...869..150M,2022ApJS..259...18M}. 
In addition, these SMBHs might not yet be close to finished with their mass growth, as the most luminous $z\gtrsim6$ quasars are, allowing the study of more `typical' SMBH growth scenarios than in the brightest quasars \citep{2012Sci...337..544V,2021ApJ...914...36I,2022MNRAS.511.3751H}.
Unfortunately, spectroscopy for measuring intrinsically fainter quasar ($M_{1450} > -25$) SMBH masses with high accuracy is extremely challenging and time-consuming, even with 8~m~class telescopes \citep{2019ApJ...880...77O,2021ApJ...919...61O}. 

Despite being a rare situation, some high-$z$ quasars can be strong-gravitationally lensed. 
Such systems are one of the best alternatives to probe quasars with intrinsically lower mass and luminosity. 
They also enable us to investigate the quasar host galaxies in unprecedented detail, thanks to flux magnification and, very importantly, an increased effective spatial resolution by factors 2--10 \citep[e.g.,][]{2018MNRAS.476.5075S,2019ApJ...880..153Y,2021ApJ...917...99Y}.
In addition, lensed quasars are excellent targets for constraining the dark matter profile of the foreground deflectors \citep[e.g.,][]{2020MNRAS.491.6077G,2020MNRAS.492.3047H}, determining the Hubble constant based on the time-delay cosmography \citep[e.g.,][]{2017MNRAS.468.2590S,2020A&A...642A.193M}, and estimating the quasar accretion disk sizes \citep[e.g.,][]{2021A&A...647A.115C}.

Because of the large lensing optical depths attained at $z\gtrsim6$, it has been predicted for decades that between $\sim 1\%$ and up to a third of high-$z$ quasars should be strongly lensed, although the actual level is still unknown \citep[e.g.,][]{2002Natur.417..923W,2010MNRAS.405.2579O,2019ApJ...870L..12P,2022ApJ...925..169Y}.
Depending on the models, we anticipate that the number of lensed quasars to be discovered in an optimal data set should range from around three \citep[lensed fraction $\sim1$\%;][]{2022ApJ...925..169Y} to greater than 10 \citep[lensed fraction $>4$\%;][]{2019ApJ...870L..12P}.
However, only one such lensed quasar was found so far, J0439+1634 at $z=6.51$ \citep{2019ApJ...870L..11F}, aside from around three hundred unlensed quasars currently known.
A leading cause of this dramatic tension is the commonly used ``dropout'' technique to select high-$z$ quasar candidates, requiring any viable candidates to have a large abrupt drop in flux in bluer optical bandpasses to remove lower-$z$ contaminants. 
Up to now, this prevented galaxy-quasar lens systems from being selected as quasar candidates due to the deflector galaxy's optical flux contribution, as a re-examination of previous selection techniques showed \citep{2020ApJ...903...34A}.
Several studies have tried to look for signs of strong gravitational lensing among the known $z\gtrsim6$ quasars, finding no evidence thus far \citep[e.g.,][]{2020ApJ...904L..32D,2021ApJ...922L..24C,2022MNRAS.514.2855P}.
Therefore, developing a more advanced selection technique to reveal those hidden populations of lenses is crucial.

In this work, we propose a novel approach to search for $z\gtrsim6$ galaxy-quasar lenses, expanding the selection space missed by previous quasar surveys, created by combining the spectral energy distribution (SED) modeling and the convolutional neural network (CNN) classification.
Using multi-band optical images that are complemented with infrared photometric data, we validate our selection pipeline and provide new high-probability lensed quasar candidates.
The following is the structure of this paper.
We begin in Section~\ref{sec:data_preselection} by describing the data acquisition and preselection of the candidates through photometric color cuts and SED modeling. 
Then, Section~\ref{sec:cnn} presents the details on the lens finding using CNN, including the datasets used for training and assessing the networks.
After that, the classification results and lensed quasar candidates are discussed in detail in Section~\ref{sec:result}. 
Finally, we will close with a summary and our conclusions in Section~\ref{sec:conclusion}.

In every part of this paper, we make use of the $\Lambda$CDM cosmology where $\Omega_\Lambda=0.685$, $\Omega_\mathrm{m}=0.315$, and $H_0 = 67.4$~km~s$^{-1}$~Mpc$^{-1}$ \citep{2020A&A...641A...6P}.
Also, all of the reported magnitudes are in the AB system.

\section{Data and Candidates Preselection} \label{sec:data_preselection}

Our lensed quasar search consists of two parts:
(1)~preselection of the candidates based on their spectral color using multi-band photometric data and 
(2)~calculation of relative probabilities (of being a lensed quasar or contaminant) based on a convolutional neural network classification.

Discovering lensed quasar candidates utilizing photometric data requires the knowledge of their spectral color -- i.e., the addition of foreground galaxy and background quasar lights.
Here, the training datasets and spectral templates will be generated, which are optimized to find systems with small image separations, i.e., Einstein radii of $\theta_\mathrm{E} \leq 1\arcsec$.
We aim to improve the candidate's purity by limiting the selection to sources that we know are more likely to be lenses based on catalog-level photometry.
This method is one way to efficiently separate the candidates from most of the contaminants while minimizing the computational resources required.
The details for the first part of our search will be described in the following section.

\subsection{Photometric Catalog}
In this work, we use the Dark Energy Survey Data Release 2 \citep[DES;][]{2021ApJS..255...20A} as the primary catalog.
This survey is based on the optical imaging collected using the Dark Energy Camera \citep[DECam;][]{2008arXiv0810.3600H,2015AJ....150..150F} equipped at the Blanco 4~m telescope at Cerro Tololo Inter-American Observatory, Chile.
DES covers $\sim 5000 \deg^2$ of the southern Galactic cap in five bands with a median point-spread function (PSF) full-width at half maximum (FWHM) of $g$, $r$, $i$, $z$, $Y$ = $1\farcs11,\ 0\farcs95,\ 0\farcs88,\ 0\farcs83,\ 0\farcs90$.
The median co-added catalog depth is $g_\mathrm{DES} = 24.7$, $r_\mathrm{DES} = 24.4$, $i_\mathrm{DES} = 23.8$, $z_\mathrm{DES} = 23.1$, and $Y_\mathrm{DES} = 21.7$~mag for 1\farcs95 diameter aperture and signal-to-noise ratio (S/N) of 10.
By default, we use the aperture-based magnitude (\texttt{MAG\_AUTO}), where the aperture size varies depending on the extent of each source, reported in the main table of DES.
As an additional note, DES tile images have a pixel scale of 0\farcs263 and a fixed zero point of 30~mag.


We then complement our primary data with the near-infrared (NIR) and mid-infrared photometry from the Visible and Infrared Survey Telescope for Astronomy Hemisphere Survey Data Release 5 \citep[VHS;][]{2013Msngr.154...35M,2021yCat.2367....0M} and Wide-field Infrared Survey Explorer \citep[unWISE version;][]{2010AJ....140.1868W,2019ApJS..240...30S} catalogs, respectively, using a 2\arcsec\ cross-matching radius.
The VHS Petrosian $J$-magnitude (\texttt{Jpmag}) and unWISE photometric bands (W1 and W2) are extremely useful for distinguishing between high-$z$ quasars, low-$z$ galaxies, and MLT dwarfs \citep[e.g.,][]{2020ApJ...903...34A,2022AJ....163..251A}.
This complementary data is also an effective way of removing spurious sources such as diffraction spikes or moving objects, which are often found in one survey but not in the others.
Finally, we correct all photometric measurements from Galactic reddening by making use of the dust map from \texttt{Bayestar19}\footnote{\url{http://argonaut.skymaps.info/}} \citep{2019ApJS..240...30S} following the \cite{1999PASP..111...63F} extinction relation.
As a side note, we cross-match our main catalog with a list of MLT dwarfs, employing a 2\arcsec\ cross-matching radius \citep{2018ApJS..234....1B,2019MNRAS.489.5301C} and quasars \citep{2021arXiv210512985F} to keep track of the known objects.

\subsection{Simulating the Colors of Quasars} \label{subsec:quasar_simulation}

We proceed to model the target sources by generating 900 mock quasars. 
Here, we distribute them in the redshifts of $5.6 \leq z \leq 7.2$ and the rest-frame 1450\,\AA\ absolute magnitudes of $-30 \leq M_{1450} \leq -20$, following the quasar luminosity function from \cite{2018ApJ...869..150M} in the form of:
\begin{multline}
	\Phi(M_\mathrm{1450}, z) = \\
	\frac{10^{k(z-6)} \Phi_*}{10^{0.4(\alpha_\mathrm{qso}+1)(M_\mathrm{1450}-M^*)} + 10^{0.4(\beta_\mathrm{qso}+1)(M_\mathrm{1450}-M^*)}}.
\end{multline}
We adopt faint- and bright-end slopes of $\alpha_\mathrm{qso}=-1.23$ and $\beta_\mathrm{qso}=-2.73$, redshift evolution term of $k=-0.7$, break magnitude of $M^*=-24.9$, and normalization of $\Phi^*=10.9$ \citep{2018ApJ...869..150M}.

We then produce the aforementioned quasar spectra following the \cite{2016ApJ...829...33Y} prescription, implemented with the \texttt{SIMQSO}\footnote{\url{https://github.com/imcgreer/simqso}} simulation code \citep{2013ApJ...768..105M}.
The \texttt{SIMQSO} is designed to produce the synthetic quasar spectra that match $\sim60,000$ Sloan Digital Sky Survey \citep[SDSS;][]{2000AJ....120.1579Y,2011AJ....142...72E,2013AJ....145...10D,2016AJ....151...44D,2017AJ....154...28B} quasars at $2.2<z<3.5$.
The code also assumes that the quasar SEDs do not significantly evolve with redshift \citep{2019ApJ...873...35S}.
However, we note that $z\gtrsim6$ quasars frequently exhibit high ionization broad lines with more dramatic velocity shifts than other $z\lesssim5$ quasars of the similar luminosity \citep[e.g.,][]{2017ApJ...849...91M,2019MNRAS.487.3305M,2020ApJ...905...51S}.
Nonetheless, a large number of low-$z$ SDSS quasars still provide a good reference for building a high-$z$ quasar composite spectrum.

The primary component of our quasar spectral model is continuum emission constructed with the power-law function with four breaks.
At the rest-wavelength range of 1100--5700\,\AA, the corresponding continuum slope $\alpha_\nu$ follows a Gaussian distribution with a mean of $\mu(\alpha_\nu) = -1.5$.
Meanwhile, for wavelength ranges of 5700--10,850\,\AA, 10,850--22,300\,\AA, and $>$\,22,300\,\AA, we change the slopes to $\mu(\alpha_\nu) = -0.48$, $-1.74$, and $-1.17$, respectively.
We use a dispersion value of $\sigma(\alpha_\nu) = 0.3$ for each of the aforementioned slopes following \cite{2016ApJ...829...33Y}.

Next, we add a series of ultraviolet to optical Fe emissions to the spectral model based on the \cite{2001ApJS..134....1V}, \cite{2006ApJ...650...57T}, and \cite{1992ApJS...80..109B} templates for the rest-frame wavelengths of $<$\,2200\,\AA, 2200--3500\,\AA, and 3500--7500\,\AA, respectively.
After that, the emission lines from broad- and narrow-line regions are appended to the model, following the equivalent widths and line FWHM distributions of SDSS quasars.
Furthermore, the intergalactic medium (IGM) absorption by the Ly$\alpha$ forest is also implemented in the simulated spectra \citep{2010ApJ...721.1448S,2011ApJ...728...23W}.
On top of that, we add the Ly$\alpha$ damping wing model following the \citet{1998ApJ...501...15M} formalism, assuming a fixed value of 3\,Mpc proximity zone size, and a randomly drawn neutral hydrogen fraction value in the range of 0--10\% \citep[e.g.,][]{2019A&A...631A..85E,2020ApJ...900...37E,2022AJ....163..251A}.
Finally, the internal reddening was applied to each mock quasar spectrum using \cite{2000ApJ...533..682C} dust model with a randomly chosen $E(B-V)$ value in the range of $-0.02\ \mathrm{to}\ 0.14$.
We should note that the negative values of reddening are for accounting for the quasars model with a bluer continuum than encompassed by the original templates.
After that, the photometry is calculated from the simulated spectra, and the corresponding uncertainties are derived from the observed magnitude and error relations taken from each survey \citep[e.g.,][]{2016ApJ...829...33Y}.

\subsection{Modeling the Spectra of Deflector Galaxies} \label{subsec:galaxy_simulation}

\begin{figure*}[htb!]
	\centering
	\epsscale{1.17}
	\plotone{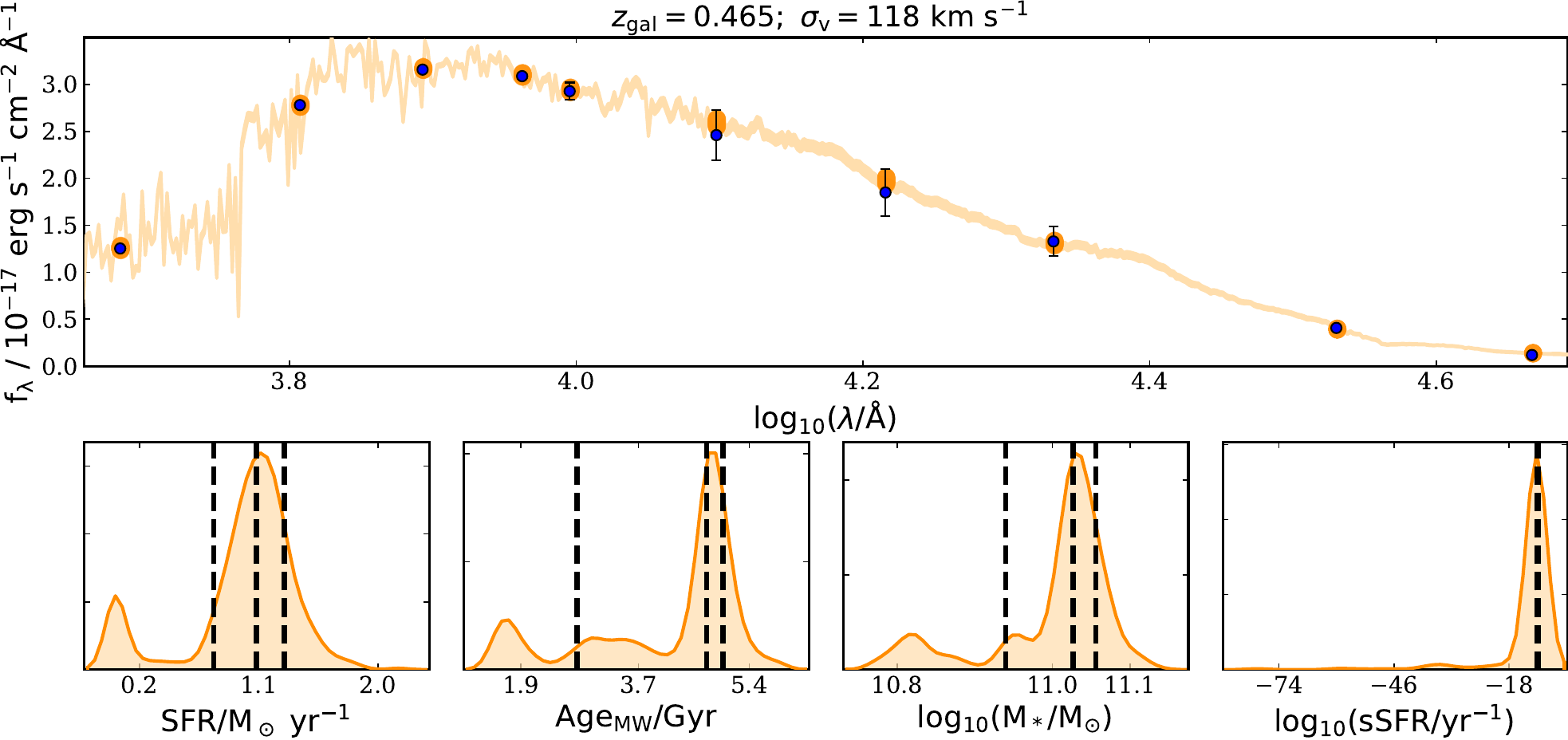}
	\caption{
		 Example of a lens galaxy fitted with the \texttt{Bagpipes} code.
		\textit{Upper panel}: The DES, VHS, and unWISE photometric data points (blue circles), posterior of the best-fit SED model (orange line), and synthesized photometry from the posterior SED (orange circles) are shown in the figure.
		\textit{Lower panel}: The posterior distribution of the derived star formation rate, mass-weighted age, stellar mass, and specific star formation rate based on the SED fitting result.
		The 16th, 50th, and 84th percentiles are marked with vertical dashed lines.
		For detailed explanations on how to calculate these quantities, we refer the reader to see \cite{2018MNRAS.480.4379C}.
	}
	\label{fig:lensgal_sedfit}
\end{figure*}

To obtain the deflectors for the lens modeling, we first search for a sample of spectroscopically confirmed galaxies using SDSS Data Release 17 \citep{2022ApJS..259...35A} catalog via the \texttt{CasJobs}\footnote{\url{http://skyserver.sdss.org/CasJobs/default.aspx}} webpage.
We retrieve all sources classified as ``GALAXY'' by the SDSS pipeline and limit our search to those having reliable velocity dispersion measurements $\sigma_v$ with less than 30\% errors, which is one important parameter to calculate the lensing effect later.
Moreover, we only select the galaxies with redshifts of $0.3 < z < 2.0$, considering that the majority of the lensing optical depth for $z\gtrsim6$ sources comes from the $z\lesssim1.5$ early-type lens galaxies. \citep{2011Natur.469..181W,2015ApJ...805...79M,2019ApJ...870L..12P,2022ApJ...925..169Y}.
After that, we cross-match these sources to the DES, VHS, and unWISE catalogs with a searching radius of 1\arcsec\ to obtain their corresponding optical/infrared magnitudes when available.
In the end, we obtain a sample of 109,808 galaxies and query their images accordingly via the \texttt{DESaccess}\footnote{\url{https://des.ncsa.illinois.edu/desaccess/}} webpage.

\begin{figure*}[htb!]
	\centering
	\epsscale{1.17}
	\plotone{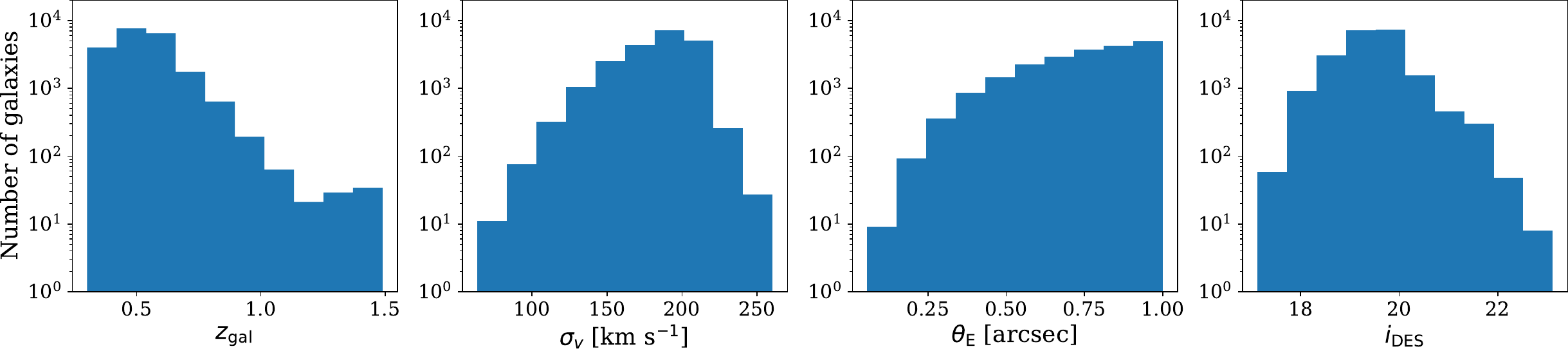}
	\caption{
		Distributions of the lens galaxy redshifts $z_\mathrm{gal}$, velocity dispersions $\sigma_v$, Einstein radii $\theta_\mathrm{E}$, and DES $i$-band magnitudes $i_\mathrm{DES}$. These parameters are used for generating the mock lenses.
	}
	\label{fig:lensgal_dist}
\end{figure*}

As the next step of our method, we need to obtain the spectra of deflectors which will be used later for constructing the mock galaxy-quasar lens SEDs.
Based on the multi-band photometric information of the aforementioned galaxies, we proceed to model their spectra using the \texttt{Bagpipes}\footnote{\url{https://bagpipes.readthedocs.io/en/latest/}} code \citep{2018MNRAS.480.4379C}. 
Several priors on the associated physical parameters need to be supplied to \texttt{Bagpipes}.
For that case, we first parametrize the star formation history (SFH) of the galaxies using a double power-law function:
\begin{equation}
	\mathrm{SFR(t) \propto \left[\left(\frac{t}{\tau}\right)^{\alpha_\mathrm{gal}} + \left(\frac{t}{\tau}\right)^{-\beta_\mathrm{gal}}  \right]^{-1}},
\end{equation}
where the falling ($\alpha_\mathrm{gal}$) and rising ($\beta_\mathrm{gal}$) slopes range between 0.01--1000 in logarithmic space \citep{2019MNRAS.490..417C}.
Then, we vary the peak times of star formation ($\tau$) between 0~Gyr and the universe's age at the observed redshifts.
Next, the prior on SFH normalization is varied in the range of $M_\mathrm{formed} = 10$--$10^{15}~M_*/M_\odot$, where $M_*$ is the stellar mass.
Also, we vary the galaxy metallicity in the range of $Z/Z_\odot=0.0$--2.5, where $Z_\odot=0.02$ is solar metallicity.

Following \cite{2019MNRAS.490..417C}, we add the \cite{2000ApJ...539..718C} dust attenuation model in the form of $A_\lambda \propto \lambda^{-n}$.
Here, we impose a Gaussian prior on the slope of the attenuation curve  $n$ with a mean and a standard deviation of 0.3 and 1.5, respectively.
The attenuation in the $V$-band $A_V$ is set to range from 0 to 8.
In addition to that, we utilize a constant value of a maximum lifetime of stellar birth-clouds of $t_\mathrm{BC}=0.01$~Gyr and a ratio of attenuation between stellar birth-clouds and the wider interstellar medium of $\epsilon = 2$.
It should be noted that the priors on the galaxy photometric redshifts are set to match the ones derived from SDSS-based spectroscopy with slight variation (i.e., $\Delta z = \pm 0.015$).
Finally, using $\sigma_v$ information from the SDSS catalog, we convolve the spectral model with a Gaussian kernel in velocity space.
As a result, we produce a sample of SEDs of galaxies ranging from ultraviolet to infrared wavelengths and their associated photometry.
Example of a lens galaxy fitted with \texttt{Bagpipes} code is presented in Figure~\ref{fig:lensgal_sedfit}.

\subsection{Constructing the Synthetic Spectra of Galaxy--Quasar Lens Systems} \label{subsec:galqso_simulation}

Finding lensed quasar candidates using catalog-level photometry requires knowledge of the shape of their SEDs -- i.e., the addition of foreground galaxy and background quasar emissions.
Here, we combine the simulated quasars and galaxies described in the previous sections to produce the mock lensed quasar spectra.

For describing the mass profile of the lenses, we assume a singular isothermal sphere \citep[SIS;][]{2015eaci.book.....S} model.
The Einstein radius can be inferred from the one-dimensional stellar velocity dispersion $\sigma_v$ in the potential of the mass distribution using the formula of:
\begin{equation}
	\theta_\mathrm{E} = 4\pi \frac{\sigma_v^2}{c^2} \frac{D_\mathrm{ds}}{D_\mathrm{s}}, 
\end{equation}
where the speed of light is $c$ while $D_\mathrm{ds}$ and $D_\mathrm{s}$ are respectively the angular diameter distances between the deflector and source and the observer and source.

It is important to note that galaxy mass distributions in nature are not perfectly symmetric.
The symmetry breaking is often caused by the ellipticity of the mass distribution or external shear forces -- e.g., the tidal gravitational field of adjacent galaxies.
Thus, this will alter the SIS lens characteristics and might produce more than two images.
However, we refrain from using a more complex lens model like Singular Isothermal Ellipsoid \citep[SIE;][]{1998ApJ...502..531B} that usually requires accurate measurements of deflector axis ratios and position angles.
Nevertheless, SIS appears to reproduce lens systems quite well, and the typical image separation remains in the same order of magnitude as predicted by the SIE model \citep{2015eaci.book.....S}.
For details on the SIS lens calculation, we refer the reader to see the Appendix~\ref{sec:sis_equation}.

\begin{figure}[htb!]
	\centering
	\epsscale{1.17}
	\plotone{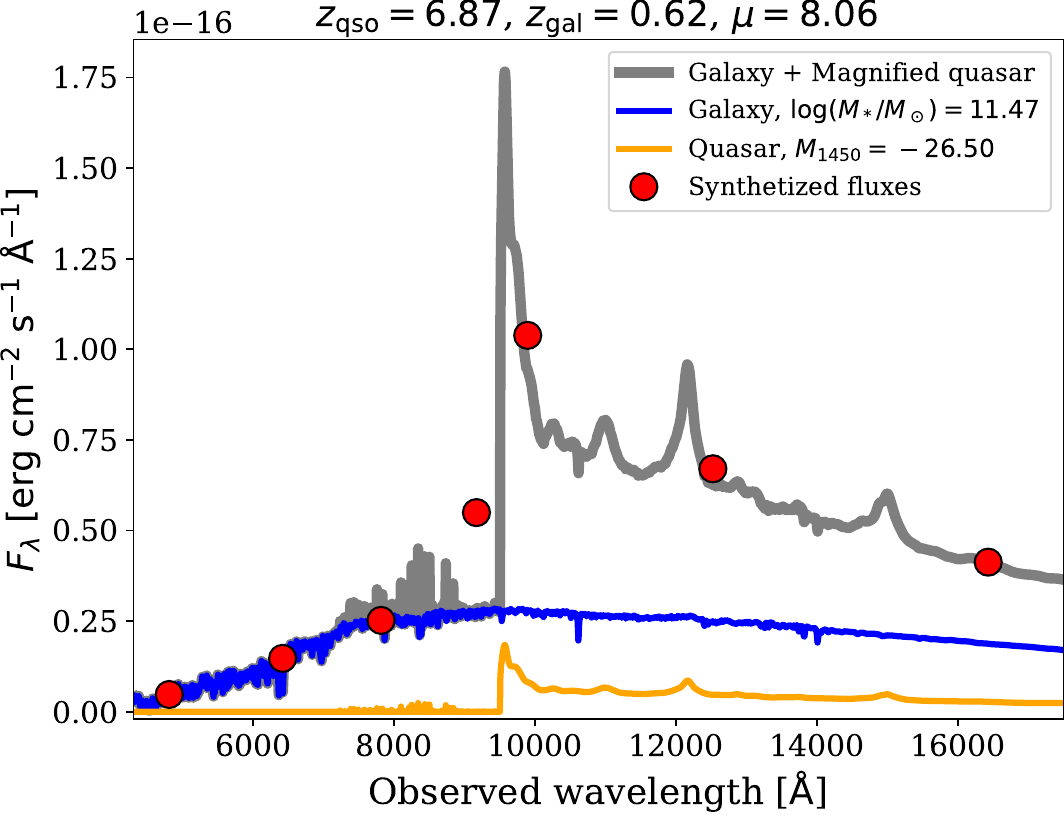}
	\caption{
		Example of the simulated galaxy-quasar lens spectrum (gray line) and the associated synthetic photometry (red circles).		
		The unlensed background quasar's emission (orange line) is magnified with a factor of $\mu$ before being added to the foreground galaxy's emission (blue line) to produce the lensed quasar spectrum.
	}
	\label{fig:qso_gal_synspec}
\end{figure}

We then produce the mock lensed quasar spectra by pairing each simulated galaxy with a randomly chosen quasar model created before.
Assuming that $\beta$ is the true angular position of the source, the quasar is then randomly placed in a specified region behind the lens within $0\farcs01 \leq \beta \leq 1\farcs5$.
After that, the source image is projected onto the lens plane, where the deflection angle and magnification are calculated based on the corresponding lens configuration.
We accept the mock lens if it contains a strong lensing effect with a magnification factor of $\mu \geq 2$ within $\theta_\mathrm{E} \leq 1\arcsec$, and the lensed quasar flux has $>5\sigma$ detection in the DES $Y$-band.
Our choice of $\beta$ and $\theta_\mathrm{E}$ ranges is motivated by \cite{2019ApJ...870L..12P}, where they predict that a significant fraction of quasars at $z\gtrsim6$ are strong-gravitationally lensed with small image separations -- i.e., $\Delta \theta = 2\theta_\mathrm{E} \leq 1 \arcsec$.

We show in Figure~\ref{fig:lensgal_dist} the distributions of lens galaxy redshifts, velocity dispersions, Einstein radii, and $i$-band magnitudes used for the simulation.
In terms of redshift, the number of galaxies is peaked at $z\approx0.5$. Then, for the $i$-band magnitudes, we see an increase of deflector numbers for up to $i_\mathrm{DES}\approx19.5$ before a rapid decrease towards the faint end.
Hence, our training dataset is skewed towards brighter, more massive lens galaxies.
This phenomenon is primarily caused by how SDSS selects the target galaxies for its spectroscopic surveys, following the criteria established by \cite{2013AJ....145...10D} and \cite{2016ApJS..224...34P}, to study the universe's large-scale structure.
Most of the target galaxies are luminous early-type galaxies at $z<1$, which are excellent probes to constrain the baryon acoustic oscillation signal and, subsequently, the universe's expansion rate \citep[e.g.,][]{2022MNRAS.511.5492Z,2022MNRAS.516.4307W}.
Note that, the faint limits for galaxies chosen for spectroscopic observations are $i=19.9$ for SDSS-III and extended to $i=21.8$ for SDSS-IV \citep{2016ApJS..224...34P}.

Finally, out of the initial 109,808 foreground galaxies and 900 background quasars, we obtain 20,823 surviving lens configurations which pass our criteria.
Naturally, the lensed quasar spectra can be obtained simply by adding the fluxes of the foreground galaxy and the magnified background quasar, as presented for one example in Figure~\ref{fig:qso_gal_synspec}.
However, in some cases, the lens galaxy could reside far enough from the quasar image, which results in two close-by objects so that the photometry extracted from the quasar image only has partial coverage of the lens galaxy flux.
To take this into account, we produce more SED templates by arbitrarily scaling the lens galaxy flux contribution in the range of 10--100\% before adding it to the magnified quasar flux.

\begin{deluxetable}{clcc}[htb!]
	\tablecaption{Summary of $z\gtrsim6$ lensed and unlensed quasar candidates selection employed in this work.}
	\label{tab:pre-selection}
	\tablehead{
		\colhead{Step} & \colhead{Selection} & \colhead{Simulated\tablenotemark{a}} & \colhead{Candidates\tablenotemark{b}}
	}	
	\startdata
	\multirow{2}{*}{1} & Initial flag criteria\tablenotemark{c} & \multirow{2}{*}{20,823} & \multirow{2}{*}{8,274,747}\\
	   & and S/N($Y_\mathrm{DES}) \geq 5$ &  & \\
	\tableline
	2  & $z_\mathrm{DES} - Y_\mathrm{DES} > 0.5$ & 3657 & 775,369 \\
	3  & $Y_\mathrm{DES} - J_\mathrm{VHS} < 1.0$ & 3657 & 735,110 \\
	4  & $Y_\mathrm{DES} - J_\mathrm{VHS} > -0.6$ & 3657 & 687,506 \\
	5  & $Y_\mathrm{DES} - W1 < 2.5$ & 3657 & 685,285 \\
	6  & $Y_\mathrm{DES} - W1 > -0.5$ & 3650 & 662,314 \\
	7  & SED modeling & 3650 & 24,723\\
	\tableline
	8 & CNN classification & -- & 448\\
	9 & Visual inspection & -- & 36\\
	\enddata
	\tablecomments{The $z_\mathrm{DES} - Y_\mathrm{DES}$ color cut limits the search to focus on the lens systems where the background quasars are located at $z=6.6$--7.2.}
	\tablenotetext{a}{Number of selected mock lenses.}
	\tablenotetext{b}{Number of real (lensed) quasar candidates selected from the DES, VHS, and unWISE catalogs.}
	\tablenotetext{c}{See the criteria in Equation~\ref{eq:imaflags}--\ref{eq:flags}.}
\end{deluxetable}

\subsection{Lensed Quasar Search via Spectral Energy Distribution Modeling} \label{subsec:sedfit}

As the next step of our lensed quasars search methodology, we select sources detected in the DES $Y$-band and having counterparts in the VHS $J$-band and unWISE W1-band within a searching radius of 2\arcsec.
Note that there are some cases where the magnitude values of the 
other bluer DES bands -- i.e., $g$, $r$, $i$, and $z$ -- are not available, or the sources are simply not detected in those bands -- i.e., S/N~$<3$.
In this case, we replace the catalog magnitudes with their associated $3\sigma$ upper limits, where we infer these limits based on the measured flux uncertainties reported in the DES table.
Then, we apply the following flag and color cuts:
\begin{eqnarray}
	\mathrm{DES\ \texttt{IMAFLAGS\_ISO}}\ (g,r,i,z,Y) = 0 \label{eq:imaflags} \\
	\mathrm{DES\ \texttt{FLAGS}}\ (g,r,i,z,Y) < 4 \label{eq:flags} \\
	\mathrm{S/N}(Y_\mathrm{DES}) \geq 5 \label{eq:colsel_start}\\
	z_\mathrm{DES} - Y_\mathrm{DES}  > 0.5 \\
	-0.6 < Y_\mathrm{DES} - J_\mathrm{VHS} < 1.0 \\
	-0.5 < Y_\mathrm{DES} - \mathrm{W1} < 2.5. \label{eq:colsel_end}
\end{eqnarray}
Equation~\ref{eq:imaflags} and \ref{eq:flags} are the primary flags to select well-behaved objects -- i.e., there is no problem in the  source extraction process, missing pixels, etc. -- from DES photometry catalog \citep{2021ApJS..255...20A}.
These criteria also ensure that our targets will have catalog entries in all DES bands.
On the other hand, the criteria in Equation~\ref{eq:colsel_start}--\ref{eq:colsel_end} are determined empirically based on the photometric features of mock lenses created in Section~\ref{subsec:galqso_simulation}.

\begin{figure}[htb!]
	\centering
	\epsscale{1.17}
	\plotone{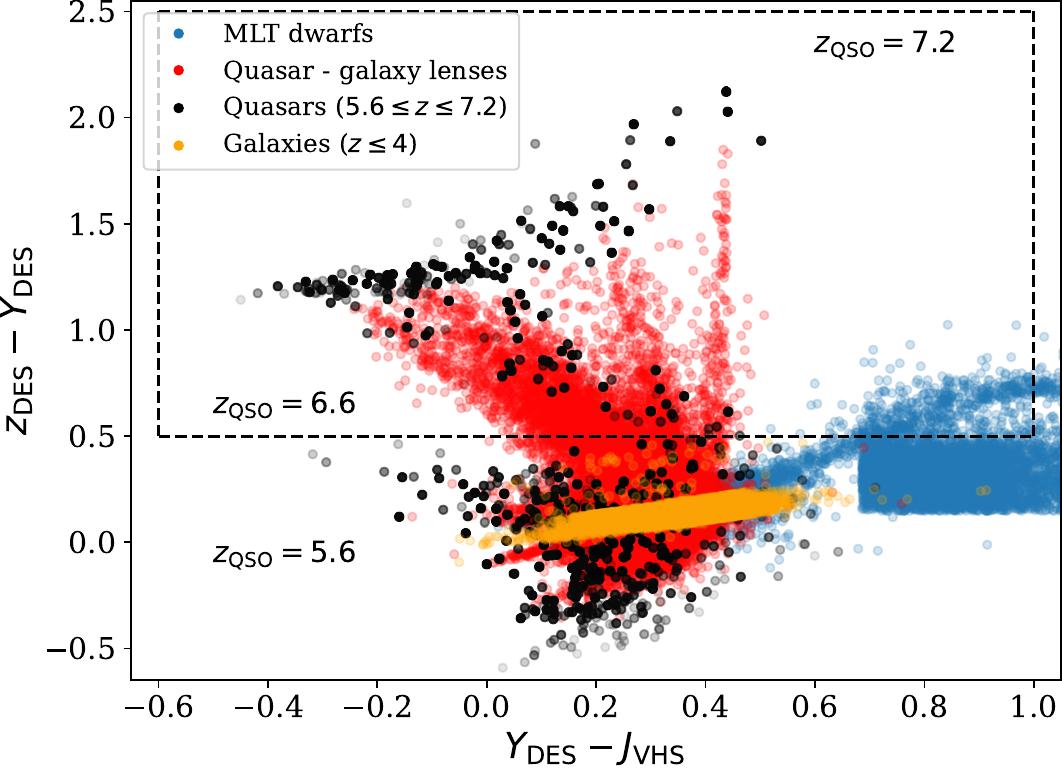}
	\caption{
		Color diagram of $Y_\mathrm{DES} - J_\mathrm{VHS}$ vs. $z_\mathrm{DES} - Y_\mathrm{DES}$.
		The colors of mock quasars (black circles), low-$z$ early-type galaxies (orange circles), simulated lenses (red circles), and a sample of MLT dwarfs \citep[blue circles;][]{2018ApJS..234....1B,2019MNRAS.489.5301C} are shown in the figure.
		Our color cuts preselection is marked by the dashed black box, which focuses on finding lensed quasars located at $z=6.6$--7.2.
		Note that the seemingly unnatural rectangle shape of the MLT dwarfs distribution on the right of the plot is caused by the color cut criteria used by \cite{2019MNRAS.489.5301C} to avoid the quasar color locus.
	}
	\label{fig:lens_color}
\end{figure}

\begin{figure*}[htb!]
	\centering
	\epsscale{1.17}
	\plotone{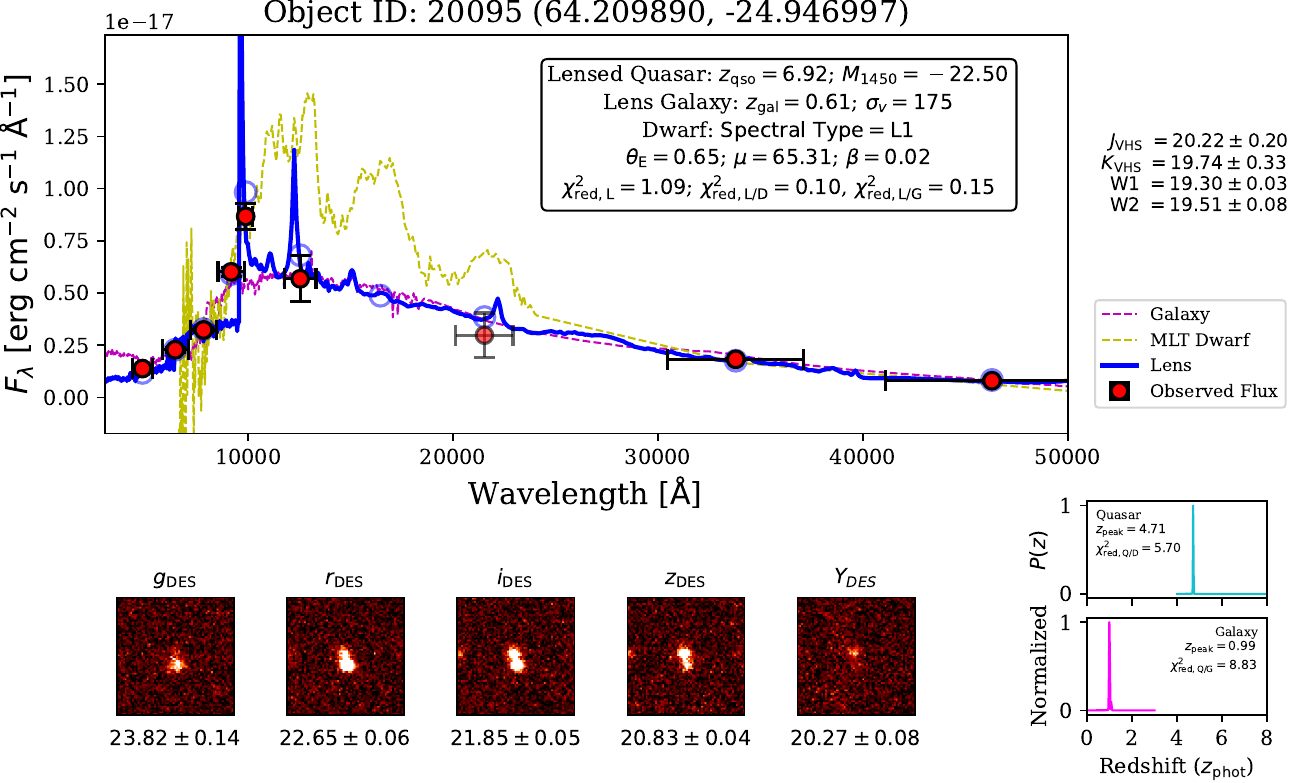}
	\caption{
		SED modeling result for a lensed quasar candidate.
		The \textit{upper panel} shows the source's observed photometry (red circles with error bars) fitted with three different templates.
		The best-fit lens spectral template is shown with a blue line, while the synthesized photometry is denoted with blue circles.
		On the other hand, best-fit models using MLT dwarf and unlensed low-$z$ galaxy templates are presented with yellow and magenta colors.
		We show the photometric redshift probability density functions in the \textit{lower right panel}, which is derived by fitting the data to unlensed high-$z$ quasar (cyan line) and low-$z$ galaxy (magenta line) templates.
		The DES cutouts with size of $16\farcs8 \times 16\farcs8$ are presented starting from the \textit{lower left panel}.
	}
	\label{fig:sed_fit}
\end{figure*}

As seen in Figure~\ref{fig:lens_color}, there is a substantial overlap between lenses and contaminants in the color space where $z_\mathrm{DES} - Y_\mathrm{DES} \leq 0.5$.
Thus, we choose to limit our color cut to $z_\mathrm{DES} - Y_\mathrm{DES} > 0.5$, which consequently focuses our search on the lens systems where the background quasars are located at $z=6.6$--7.2.
More details on the quasar selection function of our lensed quasar search criteria can be seen in Appendix~\ref{sec:color_completeness}.
To quantify our selection completeness, we start by applying $z_\mathrm{DES} - Y_\mathrm{DES} > 0.5$ color cut to the remaining 20,823 simulated lenses produced in the previous section.
As a result, only 3657 of them passed this criterion.
We then use this number of remaining mock lenses as a basis for estimating our selection completeness in recovering lens systems having the quasars at $z=6.6$--7.2.
Further cuts are made by applying criteria in Equation~\ref{eq:colsel_start}--\ref{eq:colsel_end}, which results in 3650 surviving mock lenses, or equivalent to around $99\%$ completeness.
This choice also rejects a substantial fraction ($\gtrsim70\%$) of the contaminants -- i.e., low-$z$ galaxies and MLT dwarfs.
Therefore, these are the candidates on which we will conduct SED modeling for further selection.

The next part of our preselection method is SED modeling using four different spectral templates: (i) the lensed quasars derived from the previously created mock lenses, (ii) the elliptical galaxies at $z\leq3$, (iii) the MLT dwarfs, and (iv) the $z\geq4$ unlensed quasars.
It is crucial to note that the templates (ii)--(iv) are derived empirically from observations and described in detail in Appendix~\ref{sec:empirical_templates}. 
The main goal of this SED modeling is to calculate the probability of each target being a lens or contaminant as well as its associated photometric redshift.
The SED modeling is implemented using the \texttt{EAZY} photometric redshift code \citep{2008ApJ...686.1503B}.
Here, \texttt{EAZY} will step through a grid of spectral templates, fit them to candidates' photometry, and try to find the best SED templates.
We choose as solutions the best models with the smallest reduced chi-square ($\chi^2_\mathrm{red}$), which can be calculated for each template $i$ as:
\begin{equation}
	\chi^2_{\mathrm{red}, i} = \sum_{n=1}^{N} \left(\frac{\mathrm{data}_n - f_n(\mathrm{model}_i)}{\sigma(\mathrm{data}_n)}\right)^2 \bigg/ (N - 1), 
\end{equation}
where the number of photometric data points and degree of freedom are denoted with $N$ and $(N-1)$, respectively.

The sources with a high probability of being lensed quasars are chosen based on the calculated $\chi^2_\mathrm{red}$ of the lens ($\chi^2_\mathrm{red, L}$), galaxy ($\chi^2_\mathrm{red, G}$), MLT dwarf ($\chi^2_\mathrm{red, D}$), and quasar ($\chi^2_\mathrm{red, Q}$) templates along with their associated ratios.
In addition to that, we also use the estimated quasar photometric redshift $z_\mathrm{qso}$ to also search specifically for $z\gtrsim6$ unlensed quasars.
Hence, the following criteria are employed to find high-$z$ lensed and unlensed quasar candidates:
\begin{eqnarray}
	\frac{\chi^2_\mathrm{red, L}}{\chi^2_\mathrm{red, D}} < 0.2\quad \mathrm{and} \quad \frac{\chi^2_\mathrm{red, L}}{\chi^2_\mathrm{red, G}} < 0.2, \label{eq:lens_sed}\\ 
	\nonumber \mathrm{or} \\
	z_\mathrm{qso} > 6.0\quad \mathrm{and}\quad \frac{\chi^2_\mathrm{red, Q}}{\chi^2_\mathrm{red, D}} < 0.3. \label{eq:qso_sed}
\end{eqnarray}
Here, the values in Equation~\ref{eq:lens_sed} are derived by fitting the simulated photometric data of the mock lenses created in Section~\ref{subsec:galqso_simulation}, which are optimized empirically to maximize the purity and completeness of our lensed quasar selection.
On the other hand, the criteria in Equation \ref{eq:qso_sed} are inferred empirically by modeling the SEDs of known MLT dwarfs and $z\gtrsim6$ quasars in the DES catalog \citep{2019MNRAS.489.5301C,2021arXiv210512985F}.
Finally, employing the SED fitting to the previously survived mock lenses yields 3650 remaining objects or corresponds to the completeness of around 99\% for selecting $z=6.6$--7.2 lensed quasars.
An example of a lens candidate selected using our SED modeling is presented in Figure~\ref{fig:sed_fit}, and a summary of our selection steps is reported in Table~\ref{tab:pre-selection}.

\section{Lens Finding Using Convolutional Neural Networks} \label{sec:cnn}

The second part of our lensed quasar search method is based on the supervised neural network classification, which requires realistic training data as input.
Convolutional neural network (CNN) has shown to be highly effective in pattern recognition tasks, such as detecting gravitational lenses in large datasets \citep[e.g.,][]{2019A&A...625A.119M,2020ApJ...894...78H,2022MNRAS.515.5121B,2022MNRAS.512.3464W,2022MNRAS.510..500G}.
The architecture of the CNN usually depends on the problem that needs to be solved.
Typically, it consists of images as data input, which are then processed by a series of convolutional, pooling, fully connected, and output layers.
Here, we present a CNN classifier trained to recognize lensed quasars against other non-lensed sources, and the following section will explain the details of our simulation.

\subsection{Mock Lens Images Construction} \label{subsec:mock_lens_images}

\begin{figure}[htb!]
	\centering
	\epsscale{1.17}
	\plotone{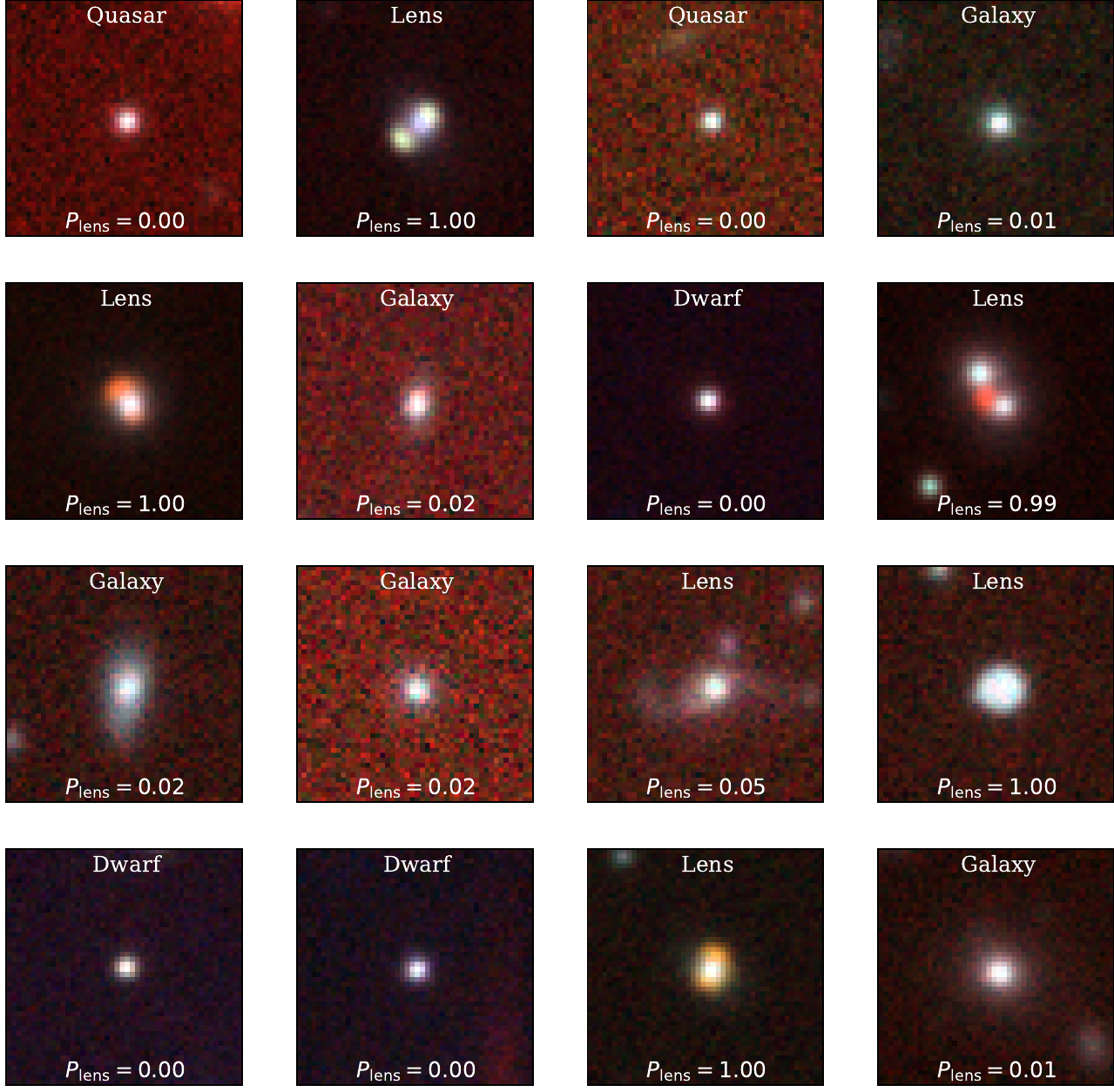}
	\caption{
		Example of negatives (galaxies and other point sources) and positive mock lenses in the training dataset.
		The cutouts are created based on $12\farcs6 \times 12\farcs6$ DES $izY$ images and stretched following the \cite{2004PASP..116..133L} arcsinh stretch algorithm to enhance contrast and improve visual appearance.
		The true labels and predicted probabilities of each source are indicated in the figure.
	}
	\label{fig:lens_train}
\end{figure}

\begin{figure*}[htb!]
	\centering
	\epsscale{1.17}
	\plotone{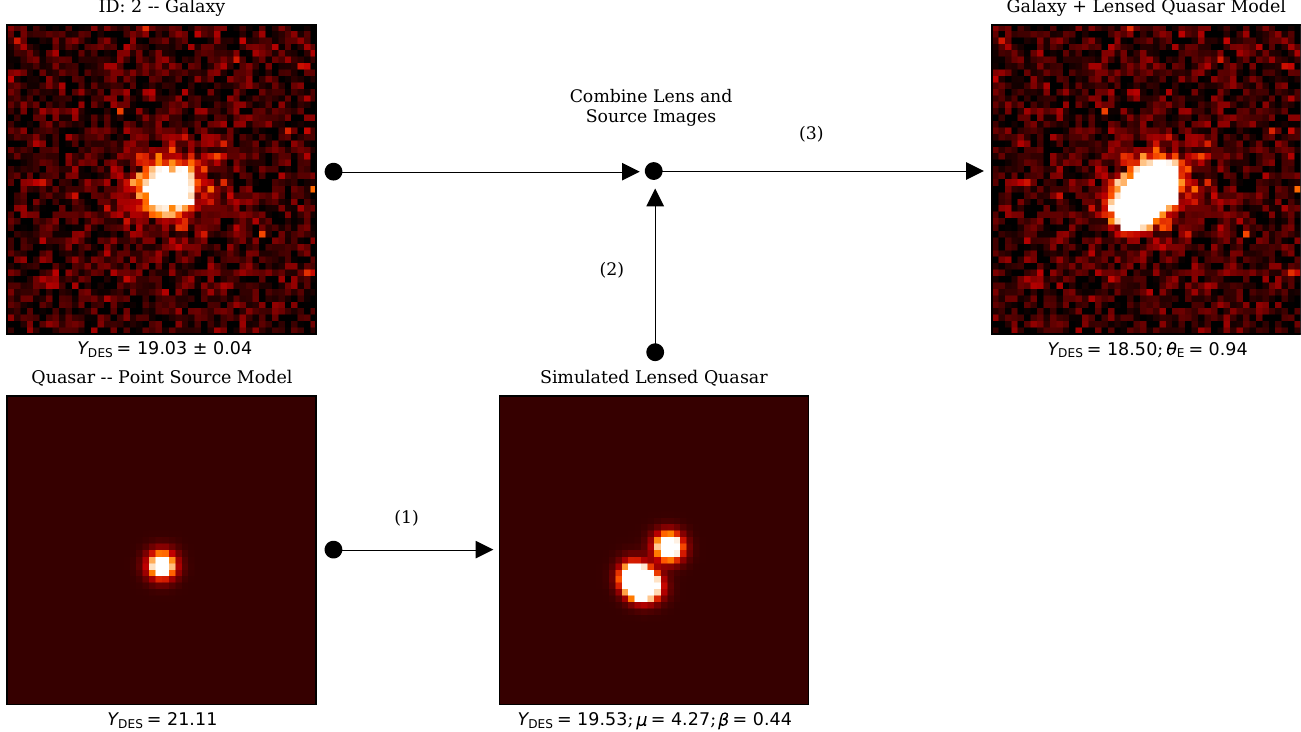}
	\caption{
		Workflow for simulating lensed quasar images. 
		The cutouts are created based on $12\farcs6 \times 12\farcs6$ DES images.
		The \textit{upper left panel} shows the actual galaxy, which we use as a lens.
		The quasar, which acts as a background source, is simulated as a point-source light convolved with a Gaussian PSF and shown in the \textit{lower left panel}.
		The \textit{lower middle panel} shows the deflected source's light, which is calculated based on the corresponding lens configuration.
		Finally, we paint the simulated arcs over an actual galaxy image and show it in the \textit{upper right panel}.
		The lens parameters (i.e., $\beta$ and $\theta_\mathrm{E}$ in arcseconds) and DES photometry are written below each panel.
		In this work, we create the mock images for all DES filters -- i.e., $grizY$-bands.
	}
	\label{fig:lens_simflow}
\end{figure*}

Since there is only one galaxy-quasar lens system found at $z\gtrsim6$ \citep{2019ApJ...870L..11F}, we need to mock up the additional lens images.
For that case, we adopt a data-driven approach for constructing the training dataset, outlined as follows.
First, we create the images of mock lenses based on the deflector galaxies, simulated quasars, and lens configurations used in Section~\ref{subsec:galqso_simulation}.
Then, through the lens equation whose parameters are defined, we painted lensed point-source lights over actual galaxy images to construct realistic galaxy-quasar lens models.
In this case, the lensing effect and ray tracing simulation are generated accordingly using \texttt{PyAutoLens}\footnote{\url{https://pyautolens.readthedocs.io/en/latest/}} lensing code \citep{2021JOSS....6.2825N}.
The produced deflected quasar lights are then convolved with a Gaussian PSF model at the location of the lens and co-added with the original DES galaxy images.
As a reminder, we accept the mock image if it contains a strong lensing case with a configuration of $0\farcs01 \leq \beta \leq 1\farcs5$ and $\mu \geq 2$ within $\theta_\mathrm{E} \leq 1\arcsec$.
For a better illustration, we present examples of mock color images created by combining $izY$-bands data in Figure~\ref{fig:lens_train} and the employed simulation workflow in Figure~\ref{fig:lens_simflow}.

\subsection{Training the Neural Networks} \label{subsec:cnn_training}

We categorize the input data for training the CNN classifier into four classes:
(i) the galaxy-quasar lenses created in the previous section,
(ii) the DES galaxies which are not selected in the lensing simulation,
(iii) a sample of MLT dwarfs that are randomly drawn from the \cite{2019MNRAS.489.5301C} catalog, 
and (iv) a sample of spectroscopically confirmed quasars at $z\sim1.5$--7.0 from the \cite{2021arXiv210512985F} database.
Class (i) and (ii) each contain 10,000 sources, while class (iii) and (iv) each consists of 8000 objects.
Hence, a total of 36,000 sources are used for the CNN input.

The images for training input are based on $grizY$-bands of DES cutouts, having the size of $48\times48$ pixels -- i.e., equivalent to angular sizes of $12\farcs6 \times 12\farcs6$.
Then, these images are min-max normalized, so the fluxes range between 0 and 1.
Note that the relative fluxes between bandpasses are conserved, so the spectral colors of the corresponding sources are preserved.
After that, we apply data augmentation to the images through random $\pm\,\pi/2$ rotations, 4-pixels translations, and horizontal or vertical flips.
This technique subsequently increases the amount of training data while enhancing the likelihood that the network will correctly classify multiple orientations of the same image.

\begin{figure*}[htb!]
	\centering
	\epsscale{1.17}
	\plotone{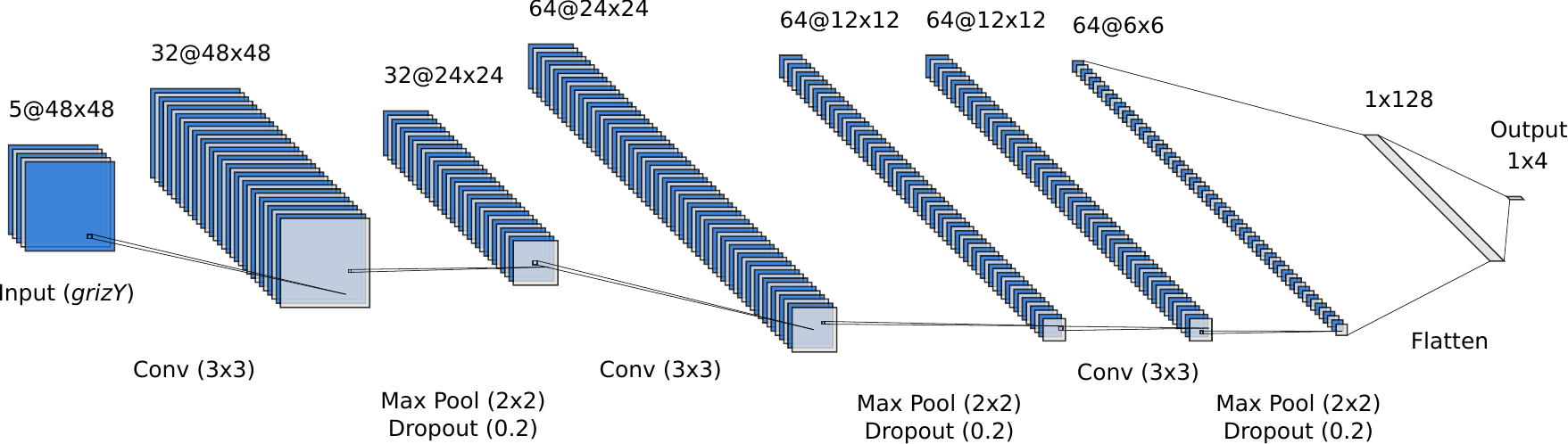}
	\caption{ 
		The architecture of the CNN used in this work.
		The input consists of DES $grizY$ images, each with a size of $48\times48$ pixels.
		The first part of our networks contains three sequences of convolutional, max pooling, and dropout layers.
		The kernel sizes and dropout rates are indicated in the figure.
		After that, the data cube will be flattened and passed through a sequence of fully connected and dropout layers before reaching the output layer.
		The softmax function in the output layer will produce four probabilities for images classification, i.e., $P_\mathrm{lens}$, $P_\mathrm{galaxy}$, $P_\mathrm{dwarf}$, and $P_\mathrm{quasar}$.  
	}
	\label{fig:cnn_arch}
\end{figure*}

Inspired by the classical CNN architectures \citep[e.g.,][]{10.1162/neco.1989.1.4.541,2019arXiv190503288S}, we start building the networks with three convolutional layers followed by one fully connected layer of 128 neurons (see Figure~\ref{fig:cnn_arch}).
The convolutional layers have a stride of $1\times1$, the ``same" padding, and a kernel size of $3 \times3 \times C$, with $C = 32,\ 64,\ 64$ for the first, second, and third layer, respectively.
For each convolutional layer, we also added a max pooling layer with a size of $2\times2 $, a stride of $2\times2$, and the same padding.
The dropout regularizations are utilized on both convolutional (drop rate~=~0.2) and fully connected layers (drop rate~=~0.5).
We also add the L2-norm regularization with a weight decay of $10^{-4}$.
The learning rate is set to $10^{-3}$ initially, while the weight and bias of each neuron are generated randomly and then updated throughout the training.
The Rectified Linear Unit (ReLu) functions are employed everywhere, except for the final layer that utilizes the softmax activation.
After passing through all of the convolutional layers, the data cube is flattened and processed by the fully connected layers to obtain the four outputs -- i.e., the probabilities of a candidate being lensed quasar, galaxy, MLT dwarf, and normal quasar.
All the training process and CNN modeling are implemented using the \texttt{TensorFlow}\footnote{\url{https://www.tensorflow.org/}} deep learning framework \citep{2016arXiv160508695A,2022zndo...4724125D}.

The specific CNN architecture mentioned above is chosen after we employ the Hyperband algorithm to optimize the model hyperparameters \citep{JMLR:v18:16-558}.
To rapidly search for a model with high performance, the Hyperband tuner applies early stopping and adaptive resource allocation method, utilizing a tournament bracket style similar to that implemented in a sports competition. 
In principle, the algorithm trains several CNN architectures with randomly-sampled configurations over a few cycles.
In other words, we explored various models with different numbers of convolutional layers, kernel sizes, and dense units.
We also tested the effect of different learning rates in the range of $10^{-2}$--$10^{-4}$, dropout rates of 0.1--0.5, convolutional filter numbers of 16--256, and L2-norm weight decays of $10^{-1}$--$10^{-5}$ to the performance of our CNN models.
Ultimately, only the top-performing half of the models are carried forward to the next round until the best model is obtained.
When compared to Bayesian optimization, random search, or grid search, this strategy maximizes the number of evaluated CNN configurations and results in more effective resource consumption.

To assess the network performance, we divide the images into three datasets: training (60\% of data), validation (20\%), and test (20\%).
Those datasets are divided further into random batches with a size of 256.
Here, we use sparse categorical crossentropy\footnote{\url{https://www.tensorflow.org/api_docs/python/tf/keras/losses/SparseCategoricalCrossentropy}} as the loss function.
For each iteration, the networks predict the outputs for one batch (forward propagation), then continue the loops through all batches from the training dataset to complete one epoch.
The information on the current batch loss is then propagated back for updating the weights and biases of the neurons, performed by following a stochastic gradient descent algorithm to minimize the loss \citep[i.e., Adam optimizer;][]{2014arXiv1412.6980K}.
This optimization is performed for each epoch, initially for all training dataset batches.
Then, the average loss for the whole training dataset is obtained.
After that, the procedures are repeated for the validation dataset but without parameter updates to the neurons. 
This way, the average loss for the validation dataset is then retrieved.
By comparing training and validation losses, we can check whether the model optimization is improved or overfitting happens.
Note that overfitting occurs when the networks fail to capture the features in the training dataset and generalize them for the unseen data.
Therefore, we randomly reorder our training data after each epoch to improve generalization and produce networks with optimum accuracy.
In the end, we can find the optimum number of training cycles by locating the epoch with the lowest average validation loss over numerous runs. 

\subsection{Evaluation of the Classifier Performance}

For each image tensor fed to our CNN model, the classifier will provide probability scores of being a lensed quasar, galaxy, MLT dwarf, or normal quasar -- i.e., $P_\mathrm{lens}$, $P_\mathrm{galaxy}$, $P_\mathrm{dwarf}$, and $P_\mathrm{quasar}$.
Then, the predicted class is then assigned by choosing which of those classes has the highest probability score.
The value of $P_\mathrm{lens} = 1$ means that there is a high chance that the classified image contains a lensed quasar.
In contrast, $P_\mathrm{lens} = 0$ means that the image is not a lensed quasar and is more similar to other contaminants.
Note that the softmax activation function that we use in the last layer of our CNN model would produce $P_\mathrm{lens} + P_\mathrm{galaxy} + P_\mathrm{dwarf} + P_\mathrm{quasar} = 1$, by construction.
In the end, our CNN training process converged after 82 epochs and obtained 96.1 (96.7) percent accuracy for the evaluation using the training (validation) dataset along with a loss value of 0.145 (0.131).

\begin{figure}[htb!]
	\centering
	\epsscale{1.17}
	\plotone{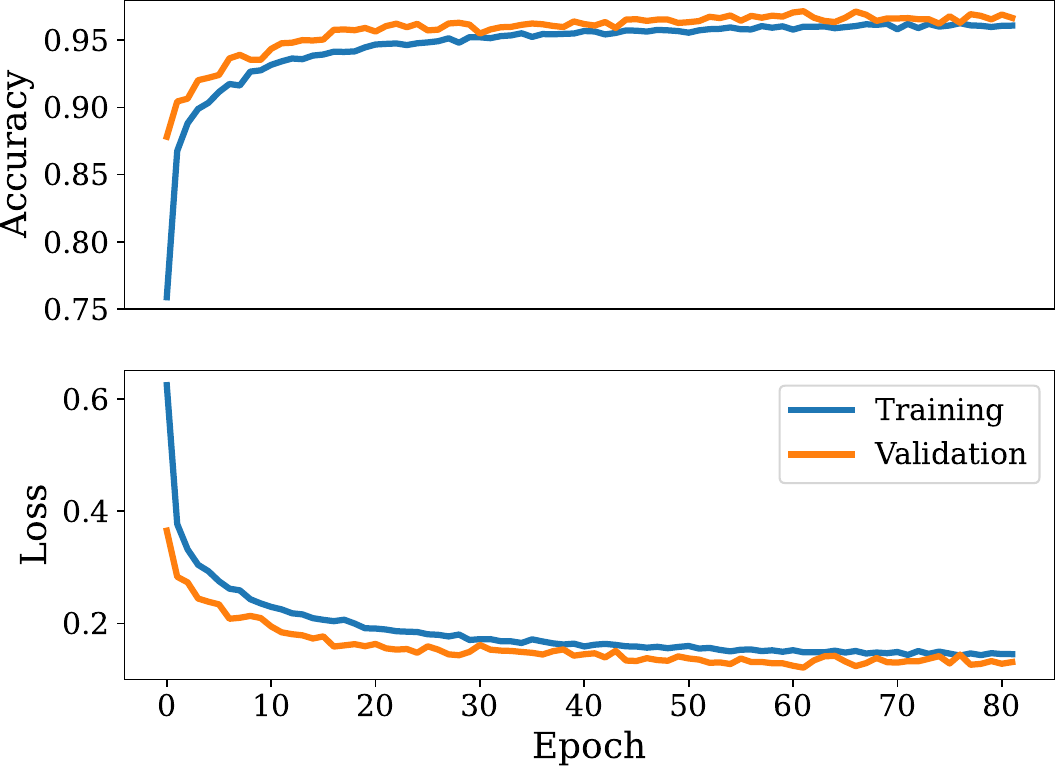}
	\caption{
		Curves of accuracy (\textit{upper panel}) and loss (\textit{lower panel}) as a function of training epoch.
		These metrics are calculated based on evaluating the CNN classifier on the training and validation datasets, which are then shown in blue and orange lines, respectively.
	}
	\label{fig:accuracy_loss}
\end{figure}

From the classical standpoint, the high accuracy attained in the training dataset might be caused by overfitting.
Thus, we compare the accuracy--loss learning curves calculated by evaluating the CNN prediction on the training and validation datasets (see Figure~\ref{fig:accuracy_loss}).
After decreasing for several epochs, the training and validation loss values stabilize and follow the same pattern.
The lack of any overfitting signals -- i.e., the training loss continues to decrease, but in contrast, the validation loss begins to increase after many epochs -- makes us confident that our CNN classifier is capable of learning and generalizing.

\begin{figure}[htb!]
	\centering
	\epsscale{1.17}
	\plotone{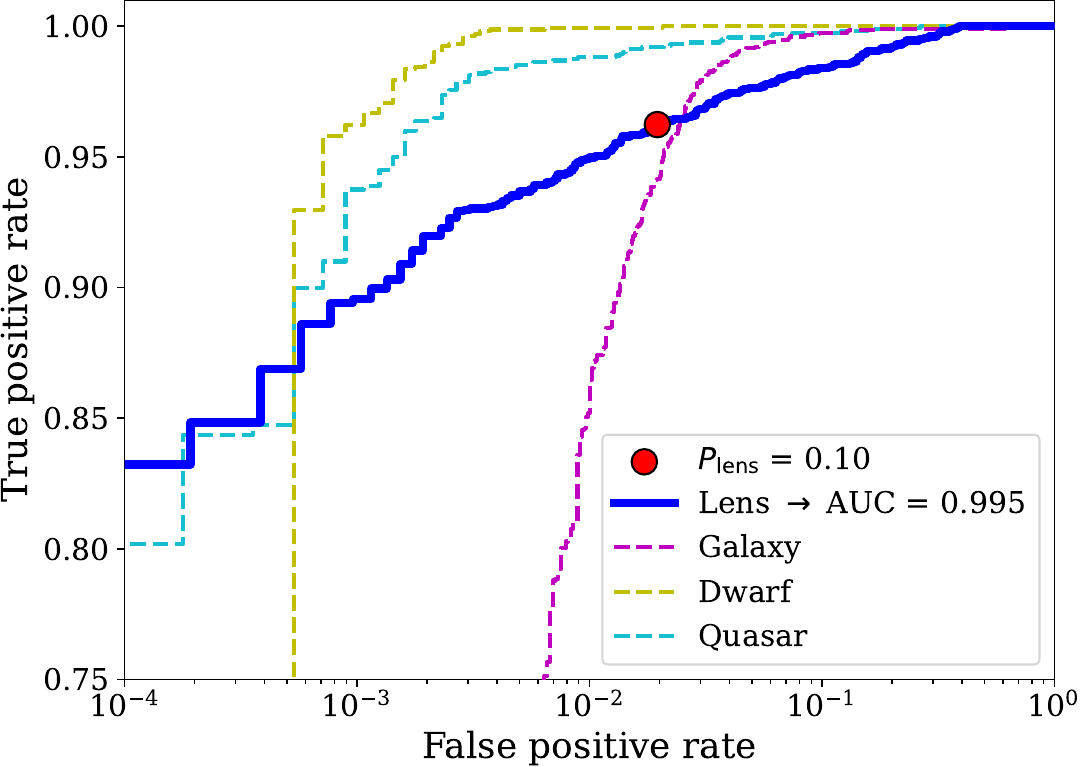}
	\caption{
		Receiver operating characteristic (ROC) curve and the associated area under the curve (AUC) value.
		The curve for classifying lensed quasars is presented with a solid blue line.
		On the other hand, the curves for predicting galaxies, MLT dwarfs, and quasars are shown with magenta, yellow, and cyan dashed lines, respectively.
		The FPR and TPR value for the adopted $P_\mathrm{lens}$ thresholds is also shown as a red circle.
	}
	\label{fig:roc_curve}
\end{figure}

Another method to assess the overall performance of the trained model is by using the receiver operating characteristic (ROC) curve and calculating its corresponding area under the curve (AUC).
In principle, it shows how well a binary classifier differentiates between two classes when the decision threshold is changed.
Thus, we first define positives (P) as the lenses and negatives (N) as the non-lenses or contaminants.
True positives (TP) are cases where the model correctly predicts the lenses, while true negatives (TN) are when the non-lenses are correctly identified.
On the other hand, false positives (FP) are where the classifier incorrectly predicts contaminants as lenses.
In addition to that, we also define false negatives (FN), the cases where the model incorrectly rejects lenses.
The ROC curve presents the false positive rate (FPR) against the true positive rate (TPR) for the validation dataset, where:
\begin{equation}
	\mathrm{
		TPR = \frac{TP}{P} = \frac{TP}{TP + FN}
	}
\end{equation}
and
\begin{equation}
	\mathrm{
		FPR = \frac{FP}{N} = \frac{FP}{FP + TN}
	}.
\end{equation}
We then construct the ROC curve by progressively raising the probability threshold from 0 to 1.
A perfect classifier will result in an AUC = 1, while a classifier that only forecasts randomly has an AUC = 0.5.
It is important to note that because we have four classes for categorizing the candidates, i.e., multi-label classification, we need to binarize our CNN output using the ``one versus all'' methodology.
Accordingly, we produce four ROC curves and present them in Figure~\ref{fig:roc_curve}, which consists of:
(i) lensed quasars against galaxies plus other point-source contaminants, shown with a solid blue line,
(ii) galaxies versus lens systems and other contaminants, shown with a dashed magenta line,
(iii) MLT dwarfs against other sources, shown with a dashed yellow line, and
(iv) quasars against lenses, galaxies, and others, shown with a dashed cyan line.
Note that we created these ROC curves based on the CNN classifier evaluation on the previously unseen test dataset, which results in high AUC values, indicating excellent performance.

Based on the ROC curves, we use the geometric mean or G-mean\footnote{The definition is G-mean~=~$\sqrt{\mathrm{TPR} \times (1-\mathrm{FPR})}$} metric to seek a balance between TPR and FPR.
The highest G-mean score indicates the best threshold of $P_\mathrm{lens}$ which maximizes TPR while minimizing FPR.
In this case, we obtain a recommended threshold of $P_\mathrm{lens} > 0.13$ which produces FPR = 0.01 and TPR = 0.96.
However, we then decide to relax this probability threshold and use $P_\mathrm{lens} > 0.1$ instead to classify the candidates as lenses, which results in FPR = 0.02 and TPR = 0.96.
This choice is made to increase the final number of lensed quasar candidates by accommodating candidates with lower lens probabilities.
Below the aforementioned $P_\mathrm{lens}$ threshold, the number of candidates will grow exponentially while their quality deteriorates, making visual examination more time-consuming and less effective.
Without a reference sample of $z\gtrsim6$ lensed quasars, it is impossible to determine the perfect threshold.
In terms of the trade-off between completeness and purity, we need to find a balance where the number of candidates is manageable for follow-up observations.

As supplementary information, we performed an additional evaluation using an independent dataset.
This test dataset is assembled based on the list of known $z\lesssim3$ lensed quasars taken from the Gravitationally Lensed Quasar Database\footnote{\url{https://research.ast.cam.ac.uk/lensedquasars/}} \citep{2012AJ....143..119I,2015MNRAS.448.1446A,2018MNRAS.475.2086A,2018MNRAS.480.1163S,2018MNRAS.479.5060L,2019MNRAS.483.4242L,2020MNRAS.494.3491L,2021MNRAS.502.1487J}.
Out of 220 lenses in the database, 34 of the known lenses have DES images.
Combined with the contaminants generated in Section~\ref{subsec:cnn_training}, we tested our CNN classifier to this new dataset and successfully recovered 27 lenses -- i.e., equivalent to the completeness of 79\% with a purity of 11\%.
We refer the reader to see Appendix~\ref{sec:known_lens_test} for more details on the resulting probability scores distribution.
Compared to the evaluation against the mock lenses, our CNN model's performance decreases when tested against the unseen real lens systems.
This decreased performance might be caused by the uniqueness of some strong lenses not covered by our simulation. 
In particular, some false negatives in the test dataset might contain lensed arcs that are too faint to be identified, compact lenses, multiple distortions caused by substructures, or other things.
Nonetheless, our CNN classifier can generalize and achieve sufficiently high accuracy for our purpose.
We should emphasize that the point of our CNN-based classification is to do a preselection, not to assemble a final quantitatively pure or complete sample, before we go into the visual inspection process.

\section{Results and Discussion} \label{sec:result}

\begin{figure*}[htb!]
	\centering
	\epsscale{1.15}
	\plotone{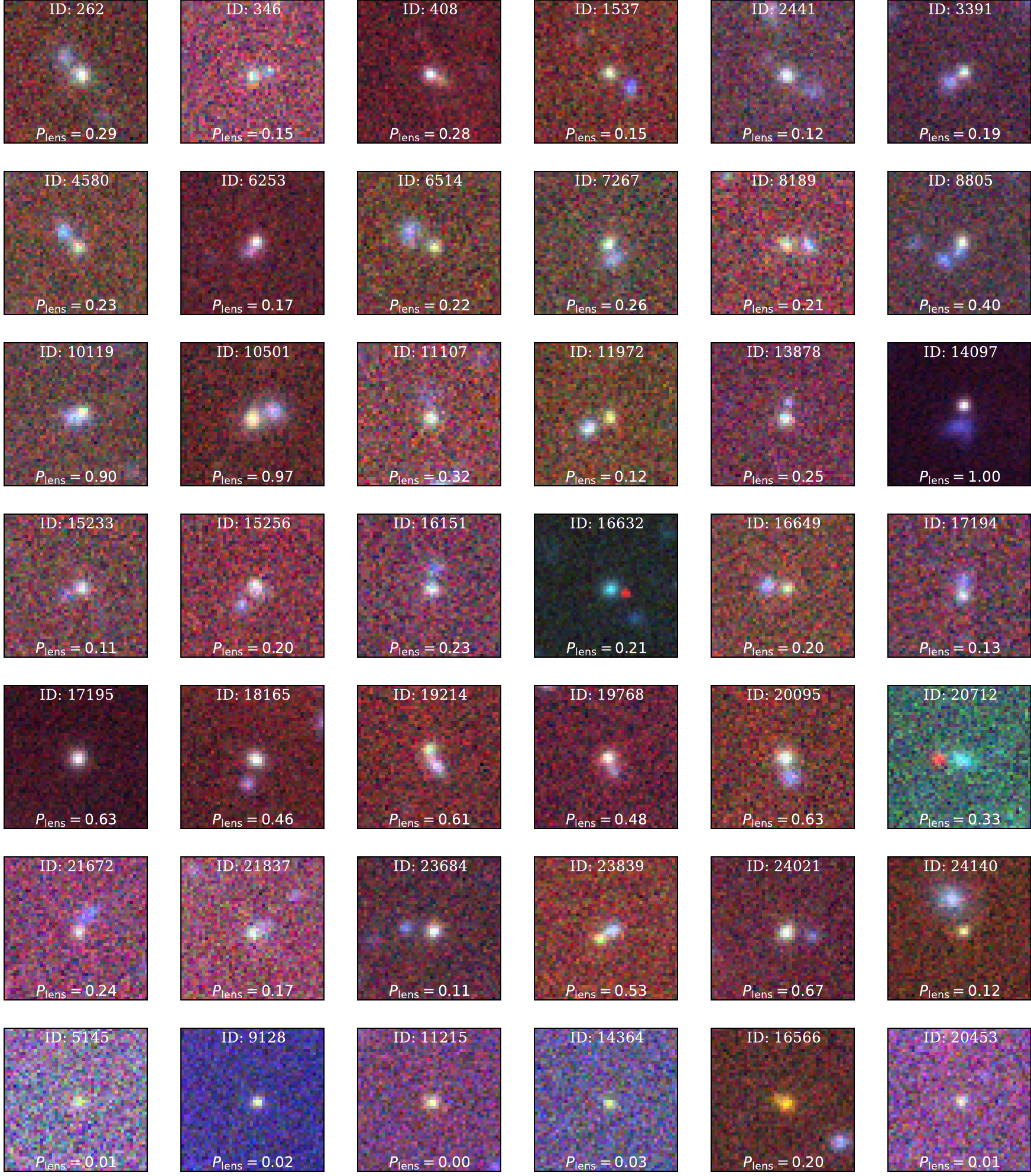}
	\caption{
		Color images of the candidates found in this paper.
		The first six rows from the top show the lens candidates, while the last row in the bottom presents the unlensed quasar candidates.
		The cutouts are created based on $12\farcs6 \times 12\farcs6$ DES $izY$ images. 
		Arcsinh stretch image normalization, following the \cite{2004PASP..116..133L} approach is applied to enhance contrast and improve visual appearance.
		The object identifiers are shown on top of each panel.
		In addition, we list the CNN-based classification scores ($P_\mathrm{lens}$) on the bottom of each cutout.
	}
	\label{fig:lens_candidates}
\end{figure*}

\begin{figure*}[htb!]
	\centering
	\epsscale{1.17}
	\plotone{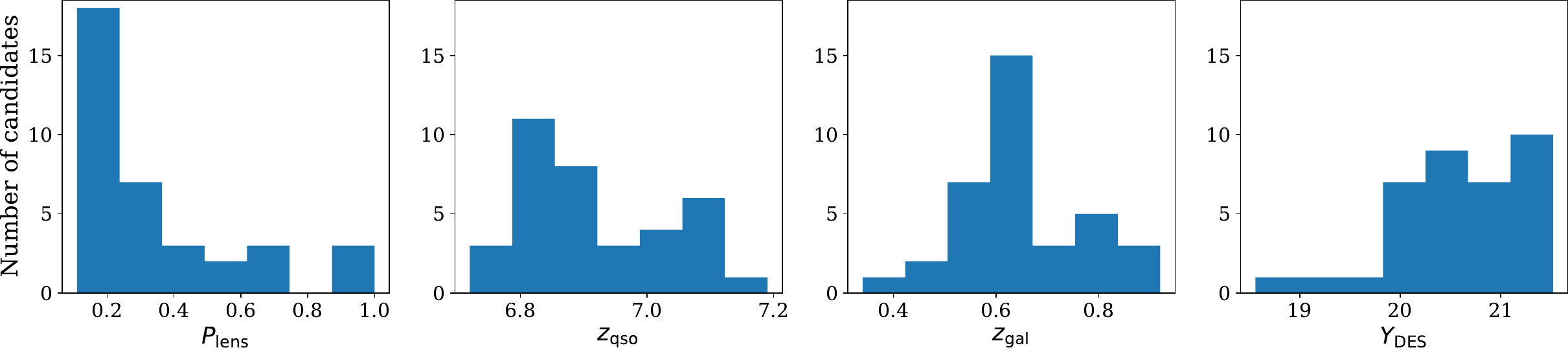}
	\caption{
		Distributions of the lens candidate CNN scores $P_\mathrm{lens}$, source redshifts $z_\mathrm{qso}$, deflector redshifts $z_\mathrm{gal}$, and DES $Y$-band magnitudes $Y_\mathrm{DES}$.
		The $z_\mathrm{qso}$ and $z_\mathrm{gal}$ parameters are inferred based on the best-fit templates from our SED modeling.
	}
	\label{fig:lenscand_dist}
\end{figure*}

We visually inspected 448 sources -- selected by our color cut, SED modeling, and CNN classifier -- which resulted in 36 high-fidelity lens candidates plus 6 unlensed quasar candidates, awaiting spectroscopic confirmation. 
More details on this selection step will be explained in the upcoming paragraphs. 
Based on the evaluation of the test dataset in Figure~\ref{fig:roc_curve}, our CNN classifier seems to have a false positive rate as low as 2\% for detecting lensed quasars at $z=6.6$--7.2.

But how many high-$z$ lensed quasars are we actually expecting to find?
The current observed lensed fraction among $z\sim6$ quasars \citep[$\approx 0.2\%$;][]{2022ApJ...925..169Y,2022AJ....163..139Y} is substantially smaller compared to previous theoretical estimations \citep[$\gtrsim4\%$;][]{2002Natur.417..923W,2019ApJ...870L..12P}.
A recent study from \cite{2022ApJ...925..169Y} argued that this tension might be caused by unaccounted biases that have not been well investigated.
Many prior investigations, in particular, relied on early measurements of quasar luminosity functions and deflector velocity dispersion functions. 
By adopting recent estimates, \citeauthor{2022ApJ...925..169Y} suggested that the lensed fraction among $z\sim6$ quasars is $\sim0.4$--0.8\% for a survey depth of $z_\mathrm{DES}=22$~mag.
Consequently, depending on assumptions, we expect between $\sim2$ \citep[lensed fraction $\approx0.6$\% of approximately three hundred known quasars;][]{2022ApJ...925..169Y} and beyond 10 \citep[lensed fraction $>4$\%;][]{2019ApJ...870L..12P} lensed quasars in an optimal final dataset.
However, we should also take into account that previous spectroscopic campaigns indicate the success rate of quasar identifications at $z\gtrsim6$ is around 20--30\% \citep[][]{2016ApJS..227...11B,2019ApJ...884...30W,2020ApJ...903...34A}.
We expect this efficiency to improve as we find more high-$z$ quasars and our training datasets expand in size.
Nonetheless, our combined SED fitting and CNN classification method is a promising way to pre-select high-$z$ lens candidates and save a lot of time before the human inspection process needs to be carried out.

Previous high-$z$ quasar searches so far found only a single lens. 
The cause of this discrepancy is rather obvious in hindsight. 
Most candidate selection approaches applied additional magnitude cuts or full `dropout'-criteria at all bandpasses bluer than the Ly$\alpha$ line \cite[e.g.,][]{2017ApJ...849...91M,2019ApJ...884...30W,2019MNRAS.487.1874R,2020ApJ...903...34A}.
This decision is understandable since the emission of $z\gtrsim6$ quasars at the wavelengths blueward of Ly$\alpha$ is strongly absorbed by the intervening IGM, creating a strong break in the spectrum and hence becoming a primary identifier for candidate preselection.
In other words, also for lensed quasars at these redshifts, we do not expect any significant flux coming from the DES $g$ or $r$ bands.
However, this is not the case for the lens galaxies at a redshift typically between $0.5<z<2$, which could contribute significant emission at $\lambda_\mathrm{obs} \lesssim 8000$\,\AA.

So, to include potential lens systems, we had to remove such dropout criteria. 
Instead, as a recap, we used the selection process as described above, hinging on downselecting candidates using spatial information that carries typical signatures of lens systems or lack thereof.
From approximately 8 million sources that existed in the combined DES, VHS, and unWISE catalogs, 662,314 of them passed our flag, S/N, and color criteria.
Then, applying the SED modeling to the previously color-selected lens candidates yields 24,723 remaining sources -- avoiding discrimination against a flux contribution by the potential lens galaxies.
After that, the CNN classifier reduces the number of candidates to only 448 sources using image color and geometry information.
Finally, visual inspection is performed on the images to discard spurious sources like moving objects, hot pixels, CCD artifacts, etc.
A summary of all selection steps that we used can be viewed in Table~\ref{tab:pre-selection}.
Also, the list of the candidates is reported in Table~\ref{tab:lens_candidates} while the corresponding composite color images and the distributions of lens candidate properties are shown in Figure~\ref{fig:lens_candidates} and Figure~\ref{fig:lenscand_dist}, respectively.

Testing our approach against past quasar candidate selection methods confirms the broadened selection space. 
If we would impose an additional cut of S/N($g_\mathrm{DES}, r_\mathrm{DES}) < 5$, or effectively the dropout criteria \citep[e.g.,][]{2016ApJS..227...11B,2019MNRAS.487.1874R,2019AJ....157..236Y}, then none of the mock lenses created in Section~\ref{subsec:galqso_simulation} would survive.
In other words, this cut would make us miss all of the lens systems containing bright galaxies as the deflectors.

\begin{figure}[htb!]
	\centering
	\epsscale{1.17}
	\plotone{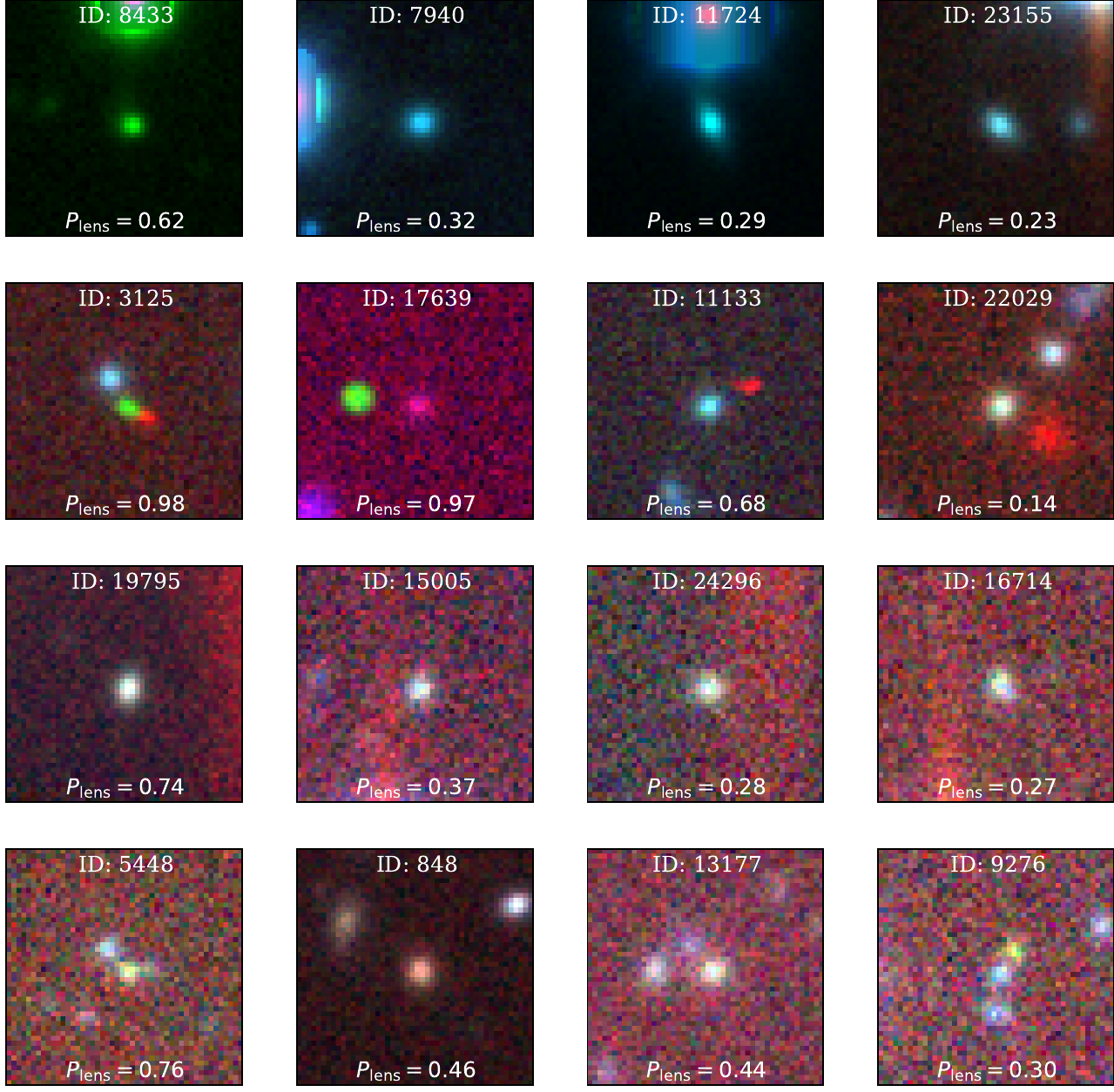}
	\caption{
		Example of the false positive sources, whose colors mimic lensed quasar SEDs, selected by our CNN classifier.
		Starting from the top rows, we show cases where the sources are located near bright stars with strong diffraction spikes, moving objects, regions with unreliable photometric data due to the presence of CCD artifacts, and candidates with less convincing strong lensing features.
		The cutouts are produced based on $12\farcs6 \times 12\farcs6$ DES $izY$ images and colorized following the \cite{2004PASP..116..133L} method.
		The predicted probabilities of each source are indicated in the figure.
	}
	\label{fig:bad_target}
\end{figure}

\begin{figure*}[htb!]
	\centering
	\epsscale{1.17}
	\plotone{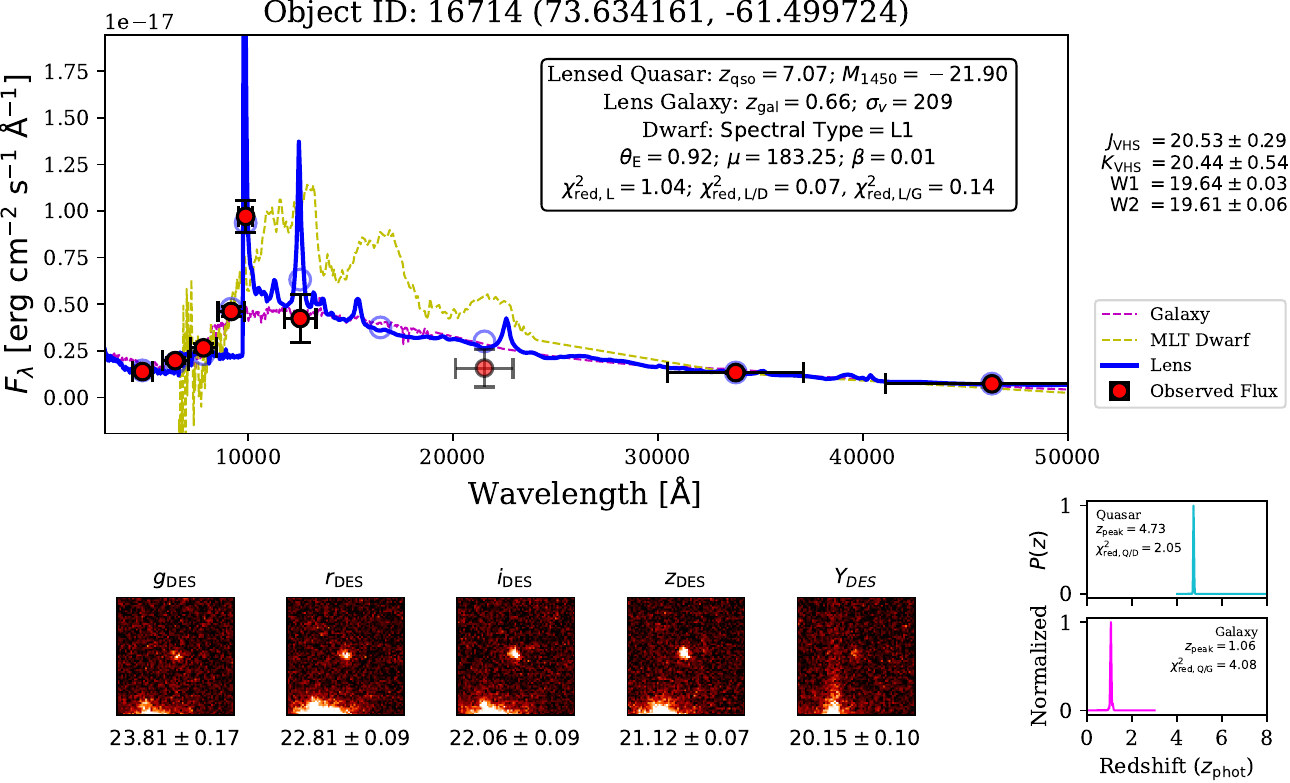}
	\caption{
		Example of the SED modeling result for a false positive candidate.
		In this case, the source is located near a bright object with strong diffraction spikes or a CCD artifact but still produces colors that mimic lensed quasar colors.
		The \textit{upper panel} displays the source's photometric data points (red circles with error bars) fitted with three distinct templates.
		The best-fit lens spectral template is presented with a blue line, while the synthesized photometry is represented with blue circles.
		On the other hand, best-fit models using MLT dwarf and unlensed low-$z$ galaxy templates are denoted with yellow and magenta colors.
		We indicate the photometric redshift probability density functions in the \textit{lower right panel}, which is inferred by fitting the data to unlensed high-$z$ quasar (cyan line) and low-$z$ galaxy (magenta line) templates.
		The DES cutouts with size of $16\farcs8 \times 16\farcs8$ are showed starting from the \textit{lower left panel}.
	}
	\label{fig:bad_sed_fit}
\end{figure*}

The crossover point to past quasar selections only comes when the deflecting galaxies are faint and consequently less massive. 
These systems make compact lenses with small Einstein radii and hence small image separations.
Subsequently, these sources are likely to have a light dominated by the quasar emission with a slightly extended shape -- at least in ground-based imaging data.
And because the images of the deflector galaxy and the lensed quasar are blended, we anticipate that neither a PSF nor a normal S\'{e}rsic profile will offer a sufficient picture of their morphology \citep[e.g.,][]{2019ApJ...870L..11F,2022AJ....163..139Y}, so using the morphological information is difficult. 
When being below some lensing mass, they would again be picked up by conventional selection methods \citep[e.g.,][]{2019MNRAS.487.1874R,2019MNRAS.484.5142P,2020ApJ...903...34A} and our SED-based selection (see Section~\ref{subsec:sedfit}), although their lensing nature would be challenging to prove. 
Of course, running this diagnostic with space-based, higher-resolution imaging data would allow for increasing the applicable parameter space substantially.

While the approach described above seems to produce sensible candidate samples, we see room for improvement. 
Currently, the rate of falsely positively classified morphologies is not zero.
However, trained astronomers can quickly dismiss the vast majority of previously unaccounted-for contaminants in this candidate list.
In principle, the false positives in our list are sources we see as highly unlikely to be lenses or having no visible signs of strong gravitational lensing features, as displayed in Figure~\ref{fig:bad_target}.
They span a wide range of apparent morphologies, for example, irregular sources or spiral arms that mimic lensing arcs.
Note that sources with highly irregular shapes that do not belong to either class in the training dataset will receive unpredictable CNN scores. 
Then, we also noted point source lights located near bright stars with strong diffraction spikes or in the regions with unreliable photometric data due to the presence of CCD artifacts, whose colors mimic lens system SEDs, as presented in Figure~\ref{fig:bad_sed_fit}.
Moreover, in some cases, we found that sources with low surface brightness in DES $Y$-band do not give sufficient information for clearly identifying lensing features.
We conclude that further research into deeper and more sophisticated CNN models may be worthwhile. 
The same is true for expansion of the simulation input: moving beyond the SIS model and adding the external shear perturbation might create subtle differences from the currently more simple models. 
In any case, the fact that we manage to expand the search parameter space by avoiding a dropout criterion while keeping the rejection rate high shows that the approach works in principle. 
We will now start to follow up on these high-probability candidates to test their quasar and lensing nature using our upcoming observations with the Focal Reducer and Low Dispersion Spectrograph 2 instrument mounted on the European Southern Observatory Very Large Telescope (110.243U.002, PI: I.T. Andika).

\section{Summary and Conclusion} \label{sec:conclusion}

In this work, we exploit the DES, VHS, and unWISE data and perform a systematic search for $z\gtrsim6$ lensed quasars. 
Our methods consist of two primary steps.
First, we pre-select the candidates based on their spectral color using catalog-level multi-band photometry, reducing the number of sources from $\approx 8$ million to just 662,314.
Second, we calculate the relative probabilities of being a lens or contaminant using CNN-based classification, resulting in 448 surviving candidates.
Note that the construction of the training dataset for CNN input is done by painting deflected point-source lights over actual DES galaxy images.
This way, we can create strong lens simulations realistically and focus on finding systems having small image separations -- i.e., Einstein radii of $\theta_\mathrm{E} \leq 1\arcsec$.
After visually inspecting the sources with a CNN score of $P_\mathrm{lens} > 0.1$, we obtain 36 newly-discovered lens candidates waiting for spectroscopic confirmation.
In addition to that, we also find 6 new unlensed quasar candidates.
These results show that automated SED modeling and CNN pipelines, supported by modest human input, are promising for detecting strong lenses from the large database.

The strategy described in this work is easily adaptable to searching lens systems at various redshifts and is very well suited for next-generation surveys such as Euclid \citep{2011arXiv1110.3193L,2022A&A...662A.112E} -- providing high-resolution imaging data over a large part of the extragalactic sky -- and the Rubin Observatory Legacy Survey of Space and Time \citep{2019ApJ...873..111I} with its very deep photometric multi-band data.
In principle, adjustments to the filter profiles, seeing distribution, and image resolution to match the target surveys are required to achieve good results.
Future searches will also benefit from extending the diversity of SEDs and morphologies for the sources and lenses in the training dataset.
Then, using more realistic galaxy mass profiles like SIE also potentially improves the classifier performance.
Moreover, experimenting with other network models like EfficientNet and ResNet might produce better results compared to the currently used classical CNN architecture \citep[e.g.,][]{2021A&A...653L...6C,2021arXiv210900014R}.
These free parameters may become more constrained when additional lenses are uncovered.
Realizing the scientific potential of our catalog of lensed quasar candidates will need spectroscopic observations for measuring the lens and source redshifts, as well as higher-resolution imaging to perform lens modeling.

\vspace{5mm}
We thank the anonymous referees for the constructive comments on the manuscript.
We would like to thank Jochen Heidt, Maarten Baes, Ilse De Looze, Archisman Ghosh, and Toon Verstraelen for the fruitful discussions and for providing valuable insights to improve this manuscript.
I.T.A. acknowledge the support from the International Max Planck Research School for Astronomy and Cosmic Physics at the University of Heidelberg (IMPRS-HD), which funded this project.
J.-T.S. acknowledges funding through the European Research Council (ERC) under the European Union's Horizon 2020 research and innovation program (grant agreement No 885301).
A.T.J. is supported by the Riset Institut Teknologi Bandung 2021.
M.O. is supported by the National Natural Science Foundation of China (12150410307). 

This project used public archival data from the Dark Energy Survey (DES). Funding for the DES Projects has been provided by the DOE and NSF (USA), MISE (Spain), STFC (UK), HEFCE (UK), NCSA (UIUC), KICP (U. Chicago), CCAPP (Ohio State), MIFPA (Texas A\&M), CNPQ, FAPERJ, FINEP (Brazil), MINECO (Spain), DFG (Germany) and the collaborating institutions in the Dark Energy Survey.
Based on observations at Cerro Tololo Inter-American Observatory, National Optical Astronomy Observatory, which is operated by the Association of Universities for Research in Astronomy (AURA) under a cooperative agreement with the National Science Foundation.

This project included data from the Sloan Digital Sky Survey (SDSS).
Funding for the SDSS-IV has been provided by the Alfred P. Sloan Foundation, the U.S. Department of Energy Office of Science, and the Participating Institutions. 
SDSS-IV acknowledges support and resources from the Center for High Performance Computing at the University of Utah. 
The SDSS website is \href{https://www.sdss.org/}{www.sdss.org}.
SDSS-IV is managed by the Astrophysical Research Consortium for the Participating Institutions of the SDSS Collaboration.

This project made use of the data obtained as part of the VISTA Hemisphere Survey, ESO Progam, 179.A-2010 (PI: McMahon).

The unWISE catalog utilized in this paper is based on data products from the Wide-field Infrared Survey Explorer, which is a joint project of the University of California, Los Angeles, and NEOWISE, which is a project of the Jet Propulsion Laboratory/California Institute of Technology.
WISE and NEOWISE are funded by the National Aeronautics and Space Administration.


\facilities{Blanco (DECam), ESO:VISTA (VIRCAM), Sloan (eBOSS/BOSS), WISE}

\software{
	Astropy \citep{2013A&A...558A..33A,2018AJ....156..123A},
	Bagpipes \citep{2018MNRAS.480.4379C},
	CosmoCalc \citep{2006PASP..118.1711W},
	EAZY \citep{2008ApJ...686.1503B},	
	Matplotlib \citep{2019zndo...2893252C},
	NumPy \citep{2020Natur.585..357H},
	Pandas \citep{2020zndo...3509134R},
	Photutils \citep{2020zndo....596036B},
	PyAutoLens \citep{2021JOSS....6.2825N},	
	SIMQSO \citep{2013ApJ...768..105M},
	SciPy \citep{2020SciPy-NMeth},
	TensorFlow \citep{2016arXiv160508695A,2022zndo...4724125D}.
}

\begin{longrotatetable}
\begin{deluxetable*}{cccccccccccccccccc}
	\tabletypesize{\scriptsize}
	\tablecaption{List of $z\gtrsim6$ lensed and unlensed quasar candidates discovered in this work.}
	\label{tab:lens_candidates}
	\tablehead{
		\colhead{ID} & 
		\colhead{Name} &
		\colhead{$g_\mathrm{DES}$} &
		\colhead{$r_\mathrm{DES}$} &
		\colhead{$i_\mathrm{DES}$} &
		\colhead{$z_\mathrm{DES}$} &
		\colhead{$Y_\mathrm{DES}$} &
		\colhead{$J_\mathrm{VHS}$} & 
		\colhead{W1} &
		\colhead{$\chi^2_\mathrm{red, Q}$} &
		\colhead{$\chi^2_\mathrm{red, G}$} &
		\colhead{$\chi^2_\mathrm{red, L}$} &
		\colhead{$\chi^2_\mathrm{red, D}$} &
		\colhead{$z_\mathrm{qso}$} &
		\colhead{$z_\mathrm{gal}$} &
		\colhead{$P_\mathrm{lens}$}	&
		\colhead{Grade}	
	}
	\startdata
	262   &  J056.71853--27.63183 &   $23.78\pm0.16$ &   $22.46\pm0.07$ &  $21.64\pm0.05$ &  $20.86\pm0.05$ &  $20.34\pm0.10$ &  $20.12\pm0.19$ &  $19.42\pm0.04$ &        37.40 &         4.57 &          0.48 &          13.90 &   6.83 &   0.67 &    0.29 &     A1 \\
	346   &  J036.96123--50.21938 &   $>25.11$ &   $23.83\pm0.16$ &  $22.78\pm0.10$ &  $21.69\pm0.08$ &  $21.06\pm0.17$ &  $20.60\pm0.34$ &  $20.03\pm0.05$ &        11.32 &         2.82 &          0.18 &           4.17 &   6.84 &   0.58 &    0.15 &     A1 \\
	408   &  J040.71110--51.68291 &   $23.14\pm0.09$ &   $22.08\pm0.05$ &  $21.17\pm0.04$ &  $20.57\pm0.04$ &  $20.06\pm0.09$ &  $19.86\pm0.17$ &  $19.10\pm0.02$ &        63.09 &         4.69 &          0.67 &          33.55 &   6.90 &   0.78 &    0.28 &     A1 \\
	1537  &  J064.46980--55.90677 &   $24.44\pm0.20$ &   $23.38\pm0.09$ &  $22.57\pm0.08$ &  $21.85\pm0.08$ &  $21.13\pm0.14$ &  $21.27\pm0.33$ &  $21.05\pm0.10$ &        21.19 &         4.05 &          0.27 &           2.40 &   6.90 &   0.62 &    0.15 &     A1 \\
	2441  &  J030.16139--23.27101 &   $24.35\pm0.23$ &   $23.02\pm0.09$ &  $22.13\pm0.07$ &  $21.17\pm0.05$ &  $20.63\pm0.09$ &  $20.27\pm0.20$ &  $19.54\pm0.04$ &        34.77 &         9.12 &          1.32 &          13.76 &   6.84 &   0.52 &    0.12 &     A1 \\
	3391  &  J092.60966--35.90861 &   $23.90\pm0.34$ &   $22.61\pm0.13$ &  $21.45\pm0.07$ &  $20.83\pm0.07$ &  $20.20\pm0.11$ &  $20.07\pm0.24$ &  $19.60\pm0.04$ &        14.82 &         3.69 &          0.34 &          14.10 &   7.09 &   0.79 &    0.19 &     A1 \\
	4580  &  J041.48368--43.88192 &   $24.70\pm0.19$ &   $23.41\pm0.08$ &  $22.87\pm0.08$ &  $21.98\pm0.06$ &  $21.31\pm0.12$ &  $20.98\pm0.25$ &  $20.33\pm0.07$ &        28.60 &         3.63 &          0.48 &           4.56 &   6.80 &   0.62 &    0.23 &     A1 \\
	6253  &  J322.05036--45.50413 &   $24.23\pm0.18$ &   $23.02\pm0.08$ &  $21.69\pm0.04$ &  $20.96\pm0.04$ &  $20.42\pm0.07$ &  $20.43\pm0.40$ &  $19.83\pm0.05$ &        37.81 &         6.00 &          0.81 &           7.75 &   6.90 &   0.85 &    0.17 &     A1 \\
	6514  &  J028.58579--52.33668 &   $25.23\pm0.32$ &   $23.81\pm0.12$ &  $22.93\pm0.10$ &  $22.11\pm0.10$ &  $21.52\pm0.18$ &  $21.27\pm0.27$ &  $20.45\pm0.07$ &        13.41 &         1.57 &          0.20 &           3.05 &   6.81 &   0.58 &    0.22 &     A1 \\
	7267  &  J025.08084--06.38508 &   $23.21\pm0.11$ &   $22.89\pm0.11$ &  $22.02\pm0.07$ &  $21.22\pm0.07$ &  $20.72\pm0.14$ &  $20.60\pm0.23$ &  $20.09\pm0.07$ &        41.18 &         3.08 &          0.49 &           6.12 &   7.00 &   0.92 &    0.26 &     A1 \\
	8189  &  J068.45382--33.48967 &   $>24.91$ &   $23.93\pm0.17$ &  $22.83\pm0.12$ &  $21.82\pm0.08$ &  $21.13\pm0.14$ &  $20.99\pm0.29$ &  $20.11\pm0.06$ &        13.84 &         1.26 &          0.09 &           3.94 &   6.84 &   0.65 &    0.21 &     A1 \\
	8805  &  J085.89481--57.30598 &   $24.19\pm0.18$ &   $22.65\pm0.05$ &  $22.10\pm0.05$ &  $21.49\pm0.05$ &  $20.97\pm0.11$ &  $20.37\pm0.21$ &  $20.85\pm0.08$ &        40.17 &         2.25 &          0.38 &           2.96 &   7.10 &   0.44 &    0.40 &     A1 \\
	10119 &  J345.07684--58.84036 &   $22.97\pm0.08$ &   $22.45\pm0.07$ &  $21.58\pm0.05$ &  $20.79\pm0.06$ &  $20.15\pm0.09$ &  $20.35\pm0.34$ &  $19.92\pm0.06$ &        50.77 &         6.58 &          0.53 &           4.16 &   6.95 &   0.78 &    0.90 &     A1 \\
	10501 &  J036.23405--40.28176 &   $24.26\pm0.26$ &   $22.03\pm0.04$ &  $20.98\pm0.03$ &  $20.24\pm0.03$ &  $19.72\pm0.07$ &  $19.07\pm0.11$ &  $18.81\pm0.02$ &        54.27 &        25.21 &          3.94 &          48.14 &   6.80 &   0.55 &    0.97 &     A1 \\
	11107 &  J042.68174--06.34957 &   $24.66\pm0.33$ &   $23.12\pm0.11$ &  $22.43\pm0.09$ &  $21.56\pm0.09$ &  $20.91\pm0.19$ &  $20.72\pm0.22$ &  $19.79\pm0.05$ &        20.99 &         2.42 &          0.39 &           5.36 &   7.06 &   0.53 &    0.32 &     A1 \\
	11972 &  J079.81120--25.37072 &   $24.94\pm0.24$ &   $23.81\pm0.11$ &  $23.24\pm0.10$ &  $22.16\pm0.08$ &  $21.18\pm0.10$ &  $21.18\pm0.32$ &  $20.44\pm0.09$ &        30.62 &         6.09 &          0.41 &           4.53 &   7.07 &   0.62 &    0.12 &     A1 \\
	13878 &  J061.07527--33.99544 &   $24.03\pm0.17$ &   $22.92\pm0.08$ &  $22.25\pm0.08$ &  $21.27\pm0.06$ &  $20.52\pm0.11$ &  $20.60\pm0.26$ &  $20.08\pm0.06$ &        24.12 &         5.07 &          0.38 &           6.91 &   6.98 &   0.60 &    0.25 &     A1 \\
	14097 &   J008.37130+01.30827 &   $22.31\pm0.08$ &   $21.13\pm0.04$ &  $20.34\pm0.03$ &  $19.45\pm0.02$ &  $18.56\pm0.04$ &  $18.98\pm0.12$ &  $18.49\pm0.02$ &       153.40 &        52.81 &          4.23 &          28.67 &   6.90 &   0.61 &    1.00 &     A1 \\
	15233 &  J025.04989--19.22226 &   $24.28\pm0.22$ &   $23.16\pm0.10$ &  $22.25\pm0.07$ &  $21.60\pm0.09$ &  $21.04\pm0.14$ &  $21.07\pm0.31$ &  $20.22\pm0.08$ &        13.85 &         1.41 &          0.15 &           9.05 &   6.90 &   0.78 &    0.11 &     A1 \\
	15256 &  J026.51434--18.52573 &   $24.79\pm0.23$ &   $23.65\pm0.11$ &  $22.59\pm0.07$ &  $21.81\pm0.06$ &  $21.29\pm0.15$ &  $21.03\pm0.33$ &  $20.75\pm0.13$ &        18.11 &         2.52 &          0.14 &           1.80 &   6.90 &   0.74 &    0.20 &     A1 \\
	16151 &  J022.21211--22.07667 &   $24.54\pm0.23$ &   $23.23\pm0.08$ &  $22.48\pm0.08$ &  $21.89\pm0.09$ &  $21.24\pm0.16$ &  $21.23\pm0.26$ &  $20.56\pm0.11$ &        16.43 &         1.26 &          0.15 &           6.72 &   7.10 &   0.61 &    0.23 &     A1 \\
	16632 &  J003.79587--43.48657 &   $24.92\pm0.35$ &   $23.11\pm0.09$ &  $21.80\pm0.04$ &  $21.00\pm0.04$ &  $20.28\pm0.05$ &  $20.20\pm0.24$ &  $19.34\pm0.03$ &        42.46 &         8.95 &          1.26 &          16.23 &   7.06 &   0.74 &    0.21 &     A1 \\
	16649 &  J004.69657--43.67283 &   $24.53\pm0.22$ &   $23.83\pm0.15$ &  $22.98\pm0.12$ &  $21.76\pm0.08$ &  $21.22\pm0.13$ &  $20.84\pm0.23$ &  $20.27\pm0.07$ &        17.12 &         3.05 &          0.37 &           3.24 &   6.78 &   0.79 &    0.20 &     A1 \\
	17194 &  J067.87033--20.39097 &   $24.00\pm0.22$ &   $22.51\pm0.08$ &  $21.88\pm0.07$ &  $21.38\pm0.09$ &  $20.78\pm0.17$ &  $20.36\pm0.18$ &  $20.23\pm0.07$ &        20.23 &         1.25 &          0.24 &           8.40 &   7.04 &   0.34 &    0.13 &     A1 \\
	17195 &  J067.92180--21.13860 &   $23.22\pm0.13$ &   $21.78\pm0.05$ &  $21.03\pm0.04$ &  $20.10\pm0.03$ &  $19.29\pm0.05$ &  $19.60\pm0.21$ &  $18.88\pm0.02$ &        76.95 &        23.61 &          1.30 &          24.89 &   7.05 &   0.67 &    0.63 &     A1 \\
	18165 &  J089.11616--24.83778 &   $23.45\pm0.15$ &   $22.08\pm0.05$ &  $21.31\pm0.05$ &  $20.61\pm0.05$ &  $20.07\pm0.09$ &  $19.85\pm0.20$ &  $19.24\pm0.03$ &        37.01 &         3.57 &          0.41 &          18.86 &   7.02 &   0.60 &    0.46 &     A1 \\
	19214 &  J331.94605--62.27649 &   $22.95\pm0.08$ &   $21.89\pm0.04$ &  $21.26\pm0.04$ &  $20.72\pm0.04$ &  $20.21\pm0.09$ &  $20.07\pm0.22$ &  $19.48\pm0.04$ &        80.55 &         3.56 &          0.44 &          21.17 &   6.77 &   0.64 &    0.61 &     A1 \\
	19768 &  J307.41039--47.88890 &   $23.72\pm0.14$ &   $22.54\pm0.07$ &  $21.52\pm0.05$ &  $20.72\pm0.04$ &  $20.10\pm0.07$ &  $20.62\pm0.44$ &  $19.20\pm0.03$ &        43.66 &         7.24 &          1.36 &          18.36 &   6.81 &   0.64 &    0.48 &     A1 \\
	20095 &  J064.20989--24.94700 &   $23.82\pm0.14$ &   $22.65\pm0.06$ &  $21.85\pm0.05$ &  $20.83\pm0.04$ &  $20.27\pm0.08$ &  $20.22\pm0.20$ &  $19.30\pm0.03$ &        63.35 &         7.18 &          1.09 &          11.12 &   6.92 &   0.61 &    0.63 &     A1 \\
	20712 &  J013.01069--00.00320 &   $>24.94$ &   $23.07\pm0.09$ &  $22.37\pm0.08$ &  $21.80\pm0.10$ &  $20.35\pm0.08$ &  $20.68\pm0.21$ &  $20.38\pm0.10$ &        25.31 &        13.21 &          2.64 &          14.90 &   7.19 &   0.44 &    0.33 &     A1 \\
	21672 &  J332.80833--43.52889 &   $>25.33$ &   $24.30\pm0.19$ &  $23.33\pm0.13$ &  $22.42\pm0.13$ &  $21.40\pm0.15$ &  $21.42\pm0.30$ &  $21.31\pm0.18$ &        10.17 &         3.65 &          0.35 &           1.78 &   6.87 &   0.69 &    0.24 &     A1 \\
	21837 &  J059.49147--51.51484 &   $23.83\pm0.13$ &   $23.15\pm0.09$ &  $22.33\pm0.08$ &  $21.65\pm0.09$ &  $20.91\pm0.15$ &  $21.05\pm0.28$ &  $20.15\pm0.05$ &        23.37 &         2.57 &          0.48 &          11.33 &   6.85 &   0.89 &    0.17 &     A1 \\
	23684 &  J060.23733--21.39838 &   $23.96\pm0.16$ &   $22.88\pm0.07$ &  $22.21\pm0.07$ &  $21.23\pm0.06$ &  $20.47\pm0.10$ &  $20.65\pm0.26$ &  $19.82\pm0.05$ &        44.95 &         6.79 &          1.07 &          10.04 &   6.81 &   0.52 &    0.11 &     A1 \\
	23839 &  J041.09428--01.94383 &   $24.03\pm0.16$ &   $22.65\pm0.06$ &  $21.77\pm0.05$ &  $21.14\pm0.05$ &  $20.50\pm0.10$ &  $20.71\pm0.33$ &  $20.31\pm0.09$ &        39.75 &         4.41 &          0.48 &           5.00 &   6.83 &   0.64 &    0.53 &     A1 \\
	24021 &  J044.34665--45.28386 &   $24.45\pm0.31$ &   $22.78\pm0.09$ &  $21.68\pm0.06$ &  $20.78\pm0.05$ &  $19.94\pm0.07$ &  $20.24\pm0.31$ &  $19.04\pm0.02$ &        45.89 &         6.05 &          1.10 &          18.41 &   6.98 &   0.67 &    0.67 &     A1 \\
	24140 &  J084.70004--19.84469 &   $25.07\pm0.33$ &   $23.54\pm0.10$ &  $22.71\pm0.08$ &  $21.73\pm0.06$ &  $21.18\pm0.14$ &  $21.00\pm0.32$ &  $20.71\pm0.11$ &        20.06 &         4.84 &          0.13 &           1.27 &   6.72 &   0.56 &    0.12 &     A1 \\
	\tablebreak
	5145  &  J051.27758--18.82994 &   $>25.47$ &   $>25.17$ &  $>24.48$ &  $22.76\pm0.17$ &  $21.48\pm0.21$ &  $21.03\pm0.25$ &  $21.05\pm0.15$ &         0.21 &         6.20 &          0.66 &           1.40 &   6.81 &   1.48 &    0.01 &     A2 \\
	9128  &  J358.77772--49.96887 &   $>24.85$ &   $>24.43$ &   $>23.88$ &  $21.22\pm0.07$ &  $20.23\pm0.07$ &  $20.21\pm0.23$ &  $20.41\pm0.09$ &         1.24 &        33.94 &          2.51 &          12.07 &   6.73 &   0.92 &    0.02 &     A2 \\
	11215 &  J080.61260--42.70559 &   $>25.47$ &    $>25.22$ &  $>24.53$ &  $21.91\pm0.06$ &  $21.09\pm0.11$ &  $20.96\pm0.28$ &  $20.10\pm0.05$ &         0.88 &        16.38 &          2.26 &           3.46 &   6.81 &   1.48 &    0.00 &     A2 \\
	14364 &  J002.72340--59.61451 &    $>25.81$ &  $>25.56$ &   $>24.78$ &  $22.76\pm0.11$ &  $22.15\pm0.20$ &  $21.61\pm0.35$ &  $21.07\pm0.12$ &         0.10 &         7.98 &          1.19 &           0.64 &   6.72 &   0.86 &    0.03 &     A2 \\
	16566 &  J070.86138--20.09059 &   $>25.38$ &   $>25.09$ &  $>24.45$ &  $21.37\pm0.05$ &  $20.65\pm0.08$ &  $20.13\pm0.21$ &  $19.75\pm0.05$ &         0.46 &        32.69 &          3.07 &           4.87 &   6.72 &   1.15 &    0.20 &     A2 \\
	20453 &  J080.68560--33.47577 &  $>24.86$ &   $>24.55$ &  $>24.01$ &  $22.16\pm0.12$ &  $21.21\pm0.16$ &  $21.43\pm0.43$ &  $20.39\pm0.07$ &         0.87 &         4.96 &          0.21 &           3.24 &   6.80 &   0.76 &    0.01 &     A2 \\
	\enddata
	\tablecomments{
        Column (1): a unique identifier for each candidate. 
        Column (2): source name. 
        Columns (3)--(7): DES $grizY$-band magnitudes and associated 1$\sigma$ uncertainties.
        Column (8): VHS $J$-band magnitude.
        Column (9): unWISE W1-band magnitude.
        Columns (10)--(13): reduced chi-square values for the quasar (Q), galaxy (G), lens (L), and MLT dwarf (D) template models.
        Columns (14)--(15): photometric redshift estimates of the background and foreground sources, based on the best-fit SED fitting template.
        Column (16): CNN classification score.
        Column (17): grade after visual inspection.
        The reported magnitudes are in the AB system, corrected for Galactic reddening, by making use of the dust map from \cite{2019ApJS..240...30S} and following the \cite{1999PASP..111...63F} extinction relation.
        If the sources are not detected -- i.e., S/N~$<3$ -- in certain bands, we replace the catalog magnitudes with their associated 3$\sigma$ upper limits, where we infer these limits based on the measured flux uncertainties reported in the DES table.
        Lens candidates are marked with the ``A1'' grade, while unlensed quasar candidates are marked ``A2.''
        To name the sources, we adopt the ``JRRR.rrrrr+DD.ddddd'' convention, where RRR.rrrrr and +DD.ddddd are the R.A. and decl. in decimal degrees (J2000), respectively.
	}
\end{deluxetable*}
\end{longrotatetable}

\appendix

\section{Singular Isothermal Sphere Lens Equation} \label{sec:sis_equation}

The singular isothermal sphere (SIS) is a simple analytical formula to model the massive galaxy mass profiles which produce strong lenses \cite[e.g.,][and references therein]{2010ARA&A..48...87T}.
The SIS profile can be expressed as:
\begin{equation}
	\beta = \theta - \theta_\mathrm{E} \frac{\theta}{|\theta|}.
\end{equation}
Here, we define $\theta_\mathrm{E}$ as the Einstein radius, $\beta$ as the true angular position of the source, and $\theta$ as the observed position of the source on the sphere relative to the center of the lens \citep[for reference, see Figure 2.31 of][]{2015eaci.book.....S}.
If $|\beta| < \theta_\mathrm{E}$, two solutions of the lens equation exist, meaning that it will produce two images at angular positions of:
\begin{equation}
	\theta_1 = \beta + \theta_\mathrm{E},\quad
	\theta_2 = \beta - \theta_\mathrm{E}.
\end{equation}
The separation of those two images does not depend on the position of the source and can be calculated using: 
\begin{equation}
	\Delta \theta = \theta_1 - \theta_2 = 2\theta_\mathrm{E}.
\end{equation}
On the other hand, for $|\beta| > \theta_\mathrm{E}$, there will be only one image of the source that exists and is located at $\theta_1$.
Finally, the magnification ($\mu$) can be determined using the formula of:
\begin{equation}
	\mu = \frac{|\theta|}{|\theta| - \theta_\mathrm{E}}.
\end{equation}

\section{Quasar Selection Function} \label{sec:color_completeness}

As mentioned in Section~\ref{subsec:quasar_simulation}, the parent population of the quasars in our simulation follows the luminosity function of \cite{2018ApJ...869..150M}. 
They are distributed in the absolute magnitudes of $-30 \leq M_{1450} \leq -20$ and redshifts of $5.6 \leq z \leq 7.2$.
Note that without strong lensings, the data we used can only detect quasars brighter than $M_{1450} \approx -24.5$. On the other hand, the lensing effect could push this limit into a fainter regime, which depends on the magnification factor.

\begin{figure}[htb!]
	\centering
	\epsscale{1.17}
	\plotone{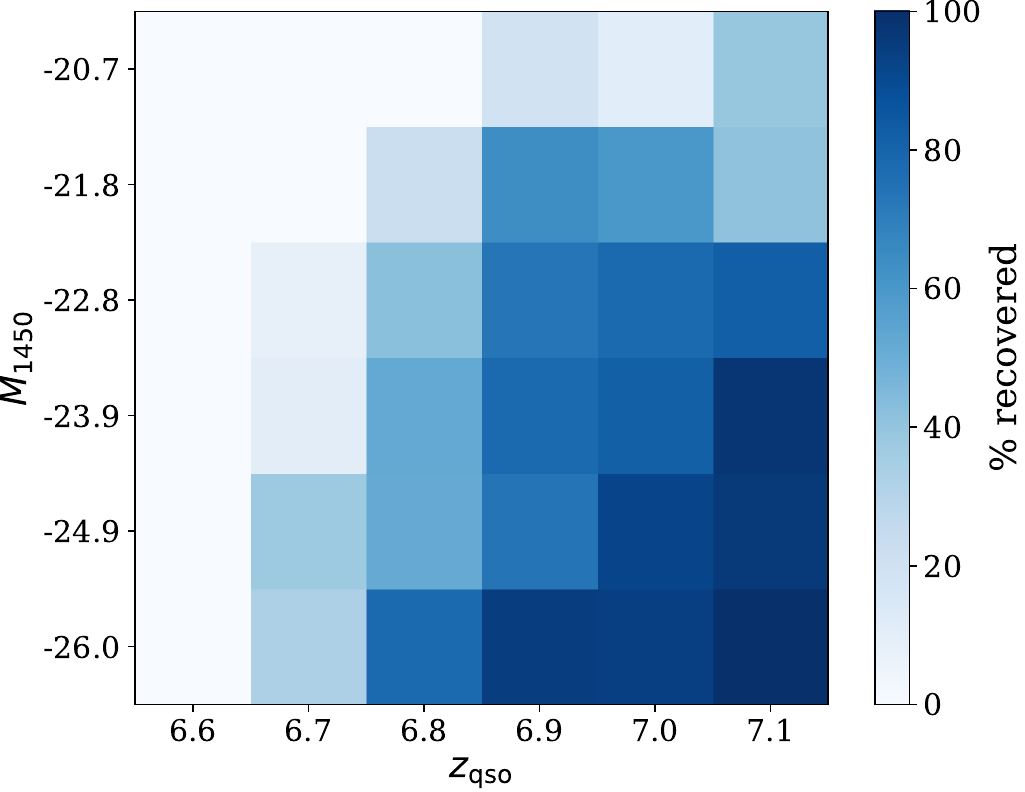}
	\caption{
		The quasar selection function that we employ in our work. 
		The probability represents the proportion of mock quasars in each ($M_{1450}$, $z$) bin that can be recovered by our classifier. 
	}
	\label{fig:color_completeness}
\end{figure}

We showed in Section~\ref{subsec:sedfit} that our selection criteria truncate the search to focus on the lens systems where the background quasars are positioned at $z=6.6$--7.2.
To see this in more detail, we first describe our selection function (or completeness) as the fraction of mock quasars with a given $M_{1450}$, $z$, and intrinsic spectral energy distribution that is recovered by our selection criteria. 
The result is presented in Figure~\ref{fig:color_completeness}.
Intrinsically faint quasars could only be detected if they are boosted by lensing with large magnification factors, which are relatively infrequent. 
Hence at a given redshift, our completeness rate drops as the quasars have intrinsically lower luminosity.

\section{Additional Spectral Templates for the Contaminants} \label{sec:empirical_templates}

To separate the candidates from contaminants, we implement the spectral energy distribution (SED) fitting method described in \cite{2020ApJ...903...34A}.
The main idea is we want to know which spectral classes best fit the observed SEDs of the candidates among the four templates -- i.e., lensed quasars, unlensed low-$z$ galaxies, MLT dwarfs, or unlensed high-$z$ quasars.

The spectral template for the MLT dwarfs are taken from SpeX Prism Library\footnote{\url{http://pono.ucsd.edu/~adam/browndwarfs/spexprism/library.html}} \citep{2014ASInC..11....7B}.
In total, there are $360$ templates that reflect stars having spectral classes of M5--M9, L0--L9, and T0--T8 in the wavelength range of 0.625--2.55 $\mu$m.
We then extrapolate these templates into the mid-infrared regime to cover unWISE bands -- i.e., W1~(3.4 $\mu$m) and W2~(4.6 $\mu$m) -- following the method presented by \cite{2020ApJ...903...34A}.

Next, we add the empirical quasar spectra taken from three sources.
The first is found in \cite{2016A&A...585A..87S}, who constructed the composite spectrum based on sample Very Large Telescope/X-shooter observations of luminous $1<z<2$ quasars obtained from the Sloan Digital Sky Survey (SDSS).
The second one is retrieved from \cite{2016ApJ...833..199J}, which is created based on 58,656 spectra of $2.1<z<3.5$ Baryon Oscillation Spectroscopic Survey (BOSS) quasars, binned by the luminosity, spectral index, and redshift.
Lastly, the third composite spectrum is taken from \cite{2016AJ....151..155H}, which is made by averaging 102,150 BOSS quasar spectra at $2.1<z<3.5$.
Note that the composites from \cite{2016AJ....151..155H} and \cite{2016ApJ...833..199J} only contain spectra with a rest-frame wavelength of up to $\sim3000$\,\AA.
To extend these further to redder wavelengths, we stitch these templates to \cite{2016A&A...585A..87S} composite spectrum starting from $2650$\,\AA.
After that, we apply the dust reddening using values of $E(B-V) = -0.02\ \mathrm{to}\ 0.14$ following the \cite{2000ApJ...533..682C} extinction law.
We further create a grid of spectra with redshift range of $4.0 \leq z \leq 8.0$ with $\Delta z = 0.003$.

In addition, to ensure that our candidates do not mimic low-$z$ unlensed elliptical galaxies, we utilize the \cite{2014ApJS..212...18B} galaxy spectral atlas.
This database contains 129 templates derived from nearby $z\lesssim0.05$ galaxies with various types (e.g., elliptical, spiral, starbursts, etc.)
We then add the reddening effect by dust using \cite{2000ApJ...533..682C} model with $A(V)=0$~to~1.
The grid of SED models is then constructed by distributing the templates across the redshifts of $0.0 \leq z \leq 3.0$ with $\Delta z = 0.005$.
Note that on top of SED models of quasars and galaxies, we add the attenuation caused by H\,\textsc{i} in the intergalactic medium using the formula from \cite{2014MNRAS.442.1805I}.

\section{Test on Known Lensed Quasars} \label{sec:known_lens_test}

We present here a sample of discovered $z\lesssim3$ lensed quasars, taken from the Gravitationally Lensed Quasar Database \citep{2012AJ....143..119I,2015MNRAS.448.1446A,2018MNRAS.475.2086A,2018MNRAS.480.1163S,2018MNRAS.479.5060L,2019MNRAS.483.4242L,2020MNRAS.494.3491L,2021MNRAS.502.1487J}, which we use for an additional independent test.
The assembly of the training data for our convolutional neural network (CNN) classifier is similar to the one explained in Section~\ref{subsec:quasar_simulation}--\ref{subsec:galqso_simulation}.
However, this time we extend the redshift of the background quasars to $1.5 \leq z \leq 7.2$ and the luminosities to $-30 \leq M_{1450} \leq -20$, and distribute the new 900 quasars uniformly in the $M_{1450}$--$z$ space.

\begin{figure}[htb!]
	\centering
	\epsscale{1.17}
	\plotone{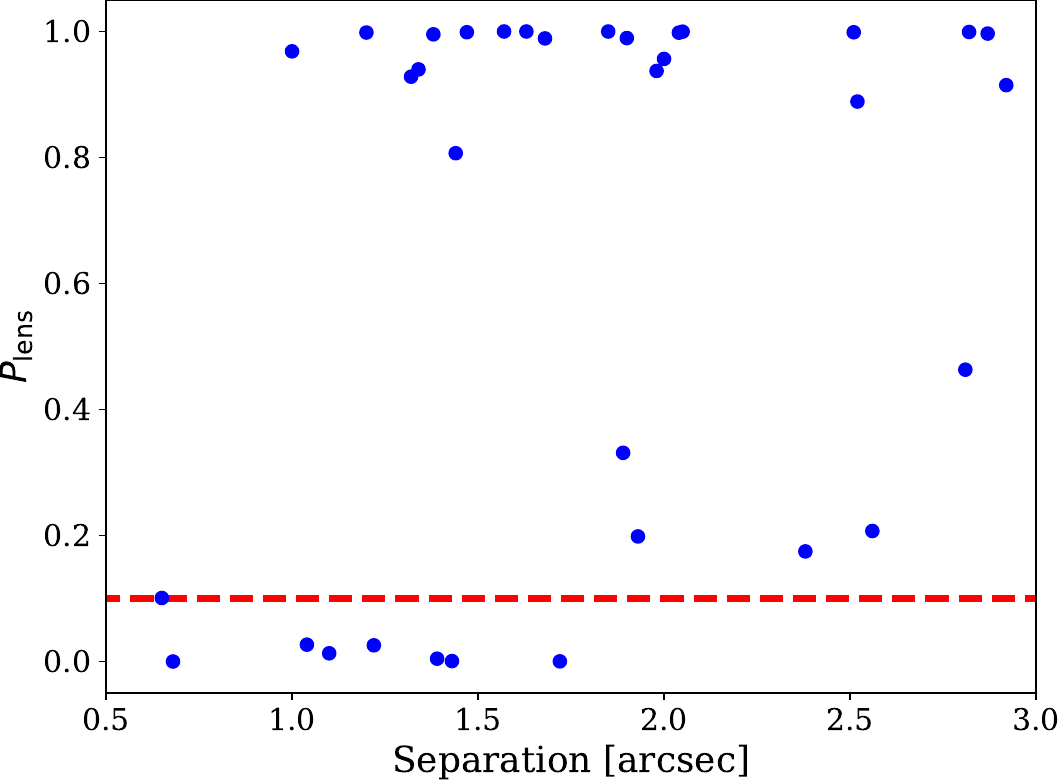}
	\\
	\plotone{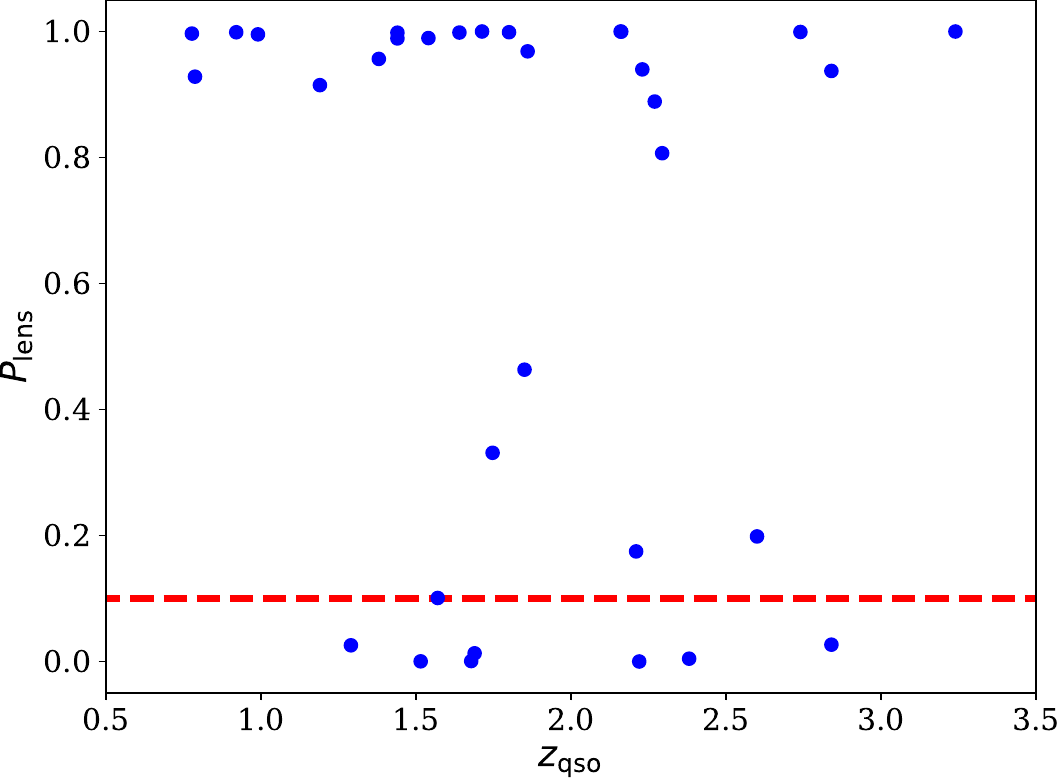}
	\caption{
		Calculated lens probabilities for a sample of known lensed quasars compared to their separations and background quasar redshifts.
		The data are shown as blue dots, while the probability threshold to classify sources as lens systems is denoted with a red dashed line.
	}
	\label{fig:CLemon_test}
\end{figure}

\begin{figure*}[htb!]
	\centering
	\epsscale{1.15}
	\plotone{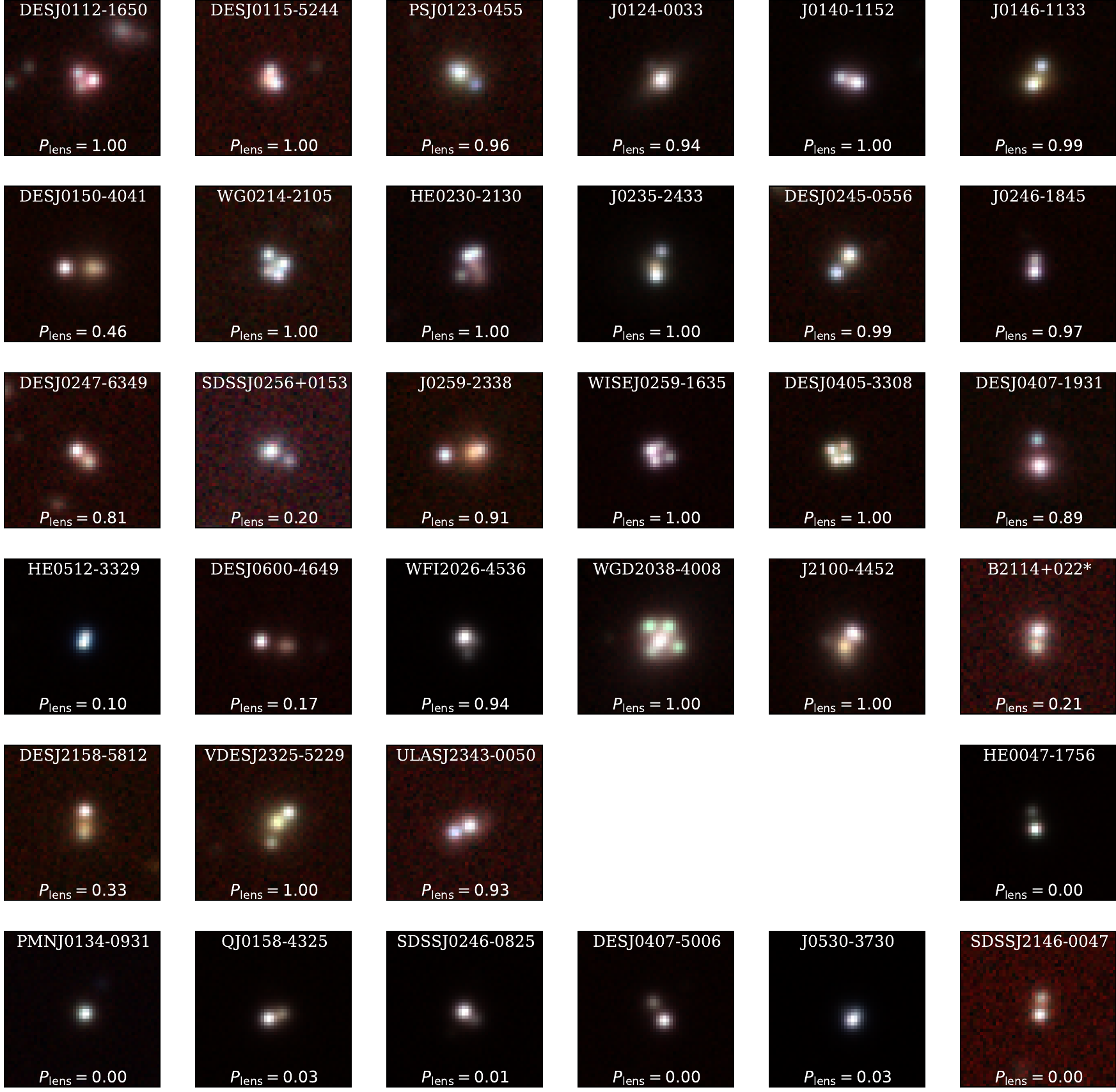}
	\caption{
		Example of known lensed quasars taken from the literature (see main text).
		The cutouts are created based on the $12\farcs6 \times 12\farcs6$ DES $izY$ images and are colorized according to the \cite{2004PASP..116..133L} method.
		The lens probabilities of the lens systems are indicated in the figure.
	}
	\label{fig:CLemon_img}
\end{figure*}

To estimate the completeness and purity of our classifier, we first combine the known lensed quasars mentioned above with a sample of contaminants -- i.e., galaxies, stars, and non-lensed quasars -- where the non-lenses are assumed to outnumber the lens population by a factor of a few thousand.
By using this test dataset, our CNN classifier successfully recovered 27 known lenses out of 34 sources available in the DES data, which is equivalent to a true positive rate (or completeness) of 79\%, with a false positive rate of 1\%.
We also found that our model has a purity of 11\% in finding the lens candidates.
The calculated lens probabilities compared to the image separations and redshifts are shown in Figure~\ref{fig:CLemon_test}.
We also show their images and lens probability scores ($P_\mathrm{lens}$) in Figure~\ref{fig:CLemon_img}.

\clearpage
\bibliography{biblio}{}
\bibliographystyle{aasjournal}

\end{document}